\documentclass[%
 reprint, 
 superscriptaddress,
 amsmath,amssymb,
 aps,
floatfix,
longbibliography
]{revtex4-1}

\usepackage{enumitem}
\usepackage{float}

\usepackage{graphicx}
\usepackage{xcolor}
\usepackage[colorlinks=true,citecolor=blue,linkcolor=magenta]{hyperref}
\usepackage{bm}

\DeclareMathOperator{\Tr}{tr}

\begin{document}


\title{Information Scrambling and Loschmidt Echo}

\author{Bin Yan}
\affiliation{Theoretical Division, Los Alamos National Laboratory, Los Alamos, New Mexico 87545}
\affiliation{Center for Nonlinear Studies, Los Alamos National Laboratory, Los Alamos, New Mexico 87545}
\affiliation{Department of Physics and Astronomy, Purdue University, West Lafayette, Indiana 47907}
\author{Lukasz Cincio}
\affiliation{Theoretical Division, Los Alamos National Laboratory, Los Alamos, New Mexico 87545}
\author{Wojciech H. Zurek}
\affiliation{Theoretical Division, Los Alamos National Laboratory, Los Alamos, New Mexico 87545}

\date{\today}

\begin{abstract}
We demonstrate analytically and verify numerically that the out-of-time order correlator is given by the thermal average of Loschmidt echo signals. This provides a direct link between the out-of-time-order correlator -- a recently suggested measure of information scrambling in quantum chaotic systems -- and the Loschmidt echo, a well-appreciated familiar diagnostic that captures the dynamical aspect of chaotic behavior in the time domain, and is accessible to experimental studies.
\end{abstract}

\maketitle

\emph{Introduction.---}
The study of the quantum version of classically chaotic systems gave rise to the field of quantum chaos, since it was realized that quantum chaotic systems share certain common characteristics \cite{Haake2006-ic,Berry2006-gh}.  In particular, in spite of the absence of the telltale exponential sensitivity to initial conditions in unitary quantum evolution, one can use as a quantum diagnostic of chaos sensitivity of the evolution to small perturbations of the Hamiltonian \cite{Peres2002-ew} or its entropy production in the presence of coupling to the environment \cite{Zurek1994-om,Zurek2006-vj}. These and related manifestations of quantum chaos have been by now widely studied \cite{Levstein1998-um,Jalabert2001-kn,Cucchietti2003-hh,Kohler2012-oz,Sanchez2016-mx,Chenu2018-sm,Chenu2019-ue,Sanchez2019-rs}, using diagnostics such as Loschmidt echo (LE) \cite{Goussev2012-ep,Gorin2006-ro},
\begin{equation}\label{eq:Loschmidt}
    M(t)\equiv |\langle \psi|e^{i H_0 t}e^{-i (H_0+V) t}|\psi \rangle|^2.
\end{equation}
This quantity incorporates the simple idea \cite{Peres2002-ew} that small perturbations of the Hamiltonian may trigger dramatic changes of the dynamics, inducing the butterfly effect.

More recently, the out-of-time-order correlator (OTOC) \cite{Larkin1969,Kitaev2015}, another diagnostic for quantum chaos, has been proposed and received considerable attention across many different fields in physics, including quantum information and high-energy and condensed matter physics \cite{Shenker2014-lz,Shenker2014-ep,Roberts2015-bo,Roberts2015-xu,Blake2016-eg,Roberts2016-nd,Maldacena2016-mb,Fu2016-dg,Hosur2016-gb,Tsuji2018-td,Swingle2017-sj,Campisi2017-ev,Chen2017-yh,Rozenbaum2017-fh,Dora2017-mf,Hashimoto2017-ug,Lin2018-mm,Lin2018-px,Von_Keyserlingk2018-cf,Nahum2018-ta,Khemani2018-cu,Gharibyan2018-fe,Xu2018-hw,Tuziemski2019-kq}. The OTOC is defined as a four-point correlator with unusual time ordering:
\begin{equation}\label{eq:otoc}
    F_\beta(t)\equiv \langle A^\dagger(t)B^\dagger A(t)B\rangle_\beta,
\end{equation}
where $A$ and $B$ are typically chosen as local operators; $A(t)=e^{iHt}Ae^{-iHt}$ is the Heisenberg operator evolving under total Hamiltonian $H$ and the average $\langle\rangle$ is taken over a thermal state at the inverse temperature $\beta$. In chaotic systems, the OTOC exhibits fast decay and converges to a persistent small value \cite{Kitaev2015}. It was argued that under certain natural assumptions, the exponential decay rate is bounded by $\lambda\le 2\pi/\beta$ \cite{Maldacena2016-mb,Tsuji2018-td}. Another benefit of OTOC is that it is designed to probe the spreading of local information over the entire system. Moreover, for systems with spatial structures, information measured by the OTOC propagates ballistically with a finite velocity known as the butterfly velocity \cite{Roberts2016-nd,Nahum2018-ta,Khemani2018-cu}. 
 
The OTOC is typically understood as an intrinsic echo-type quantity. For instance, when $A$ and $B$ are chosen as unitary operators, Eq.~(\ref{eq:otoc}) can be directly measured by echo experiments \cite{Swingle2016-hh, Garttner2017-rl, Li2017-il,Landsman2019-lf,Sanchez2019-rs}. There are efforts to build more direct links between these two quantities, e.g., using variants of the OTOC and LE or particular choices of operators for the OTOC evaluation \cite{Kurchan2018-ff,Schmitt2019-gg,Hamazaki2018-ep}.
However, the precise quantitative equivalence between the OTOC and the Loschmidt echo is still missing. Establishing such a relation would be beneficial for both areas and shed new light on the whole field of quantum chaos. 

In this work, we accomplish the task of connecting the OTOC to the Loschmidt echo. We shall focus on the temporal decay of the OTOC without extra complications caused by spatial propagation. We demonstrate that the OTOC equals the thermal average of the Loschmidt echo. The coupling between the target local systems, i.e., the supports of the local operators, and the rest of the total system plays the role of a perturbation. To further support our theory, we present two model studies involving coupled inverted harmonic oscillators and a random matrix model.

\emph{Bridging the out-of-time-order correlator and Loschmidt echo.---}For a chaotic Hamiltonian, the universal decay of the OTOC is insensitive to the form of operators $A$ and $B$ in Eq.~(\ref{eq:otoc}), as long as they are generic, i.e., not reflecting the particular symmetries possessed by the Hamiltonian. Any generic choice of local operators, e.g., random operators, is representative for the universal decay of the OTOC. This allows us to look at the typical behavior of the OTOC by averaging all unitary operators on subsystem $S_A$ and $S_B$:
\begin{equation}
    \overline{F_\beta(t)}\equiv \int dAdB\ F_\beta(t),
\end{equation}
where the integration is performed with respect to the Haar measure for unitary operators. Similar ideas have been considered in the literature \cite{Fan2017-vy,Cotler2017-oi,De_Mello_Koch2019-ez,Ma2019-sw}. Here, we will assume that $A$ and $B$ are supported on distinct local subsystems. For global operators the OTOC has been shown to be closely related to the spectral form factor of the Hamiltonian \cite{Cotler2017-oi,Cotler2017-dq,De_Mello_Koch2019-ez}. As will be seen in the following, taking into account the local structure of the system is crucial to reveal the correct behavior of the OTOC. 

For simplicity, we focus on the OTOC at an infinite temperature ($\beta=0$). It is straightforward to generalize to a finite temperature by distributing the thermal density operator over a thermal loop, e.g., using the scheme described in Refs.~\cite{Maldacena2016-mb,Cotler2017-oi,Tsuji2018-td}. The finite-temperature correction will be taken into account when discussing the temperature dependence of the decay rate. 

\begin{figure}
\includegraphics[width=\columnwidth]{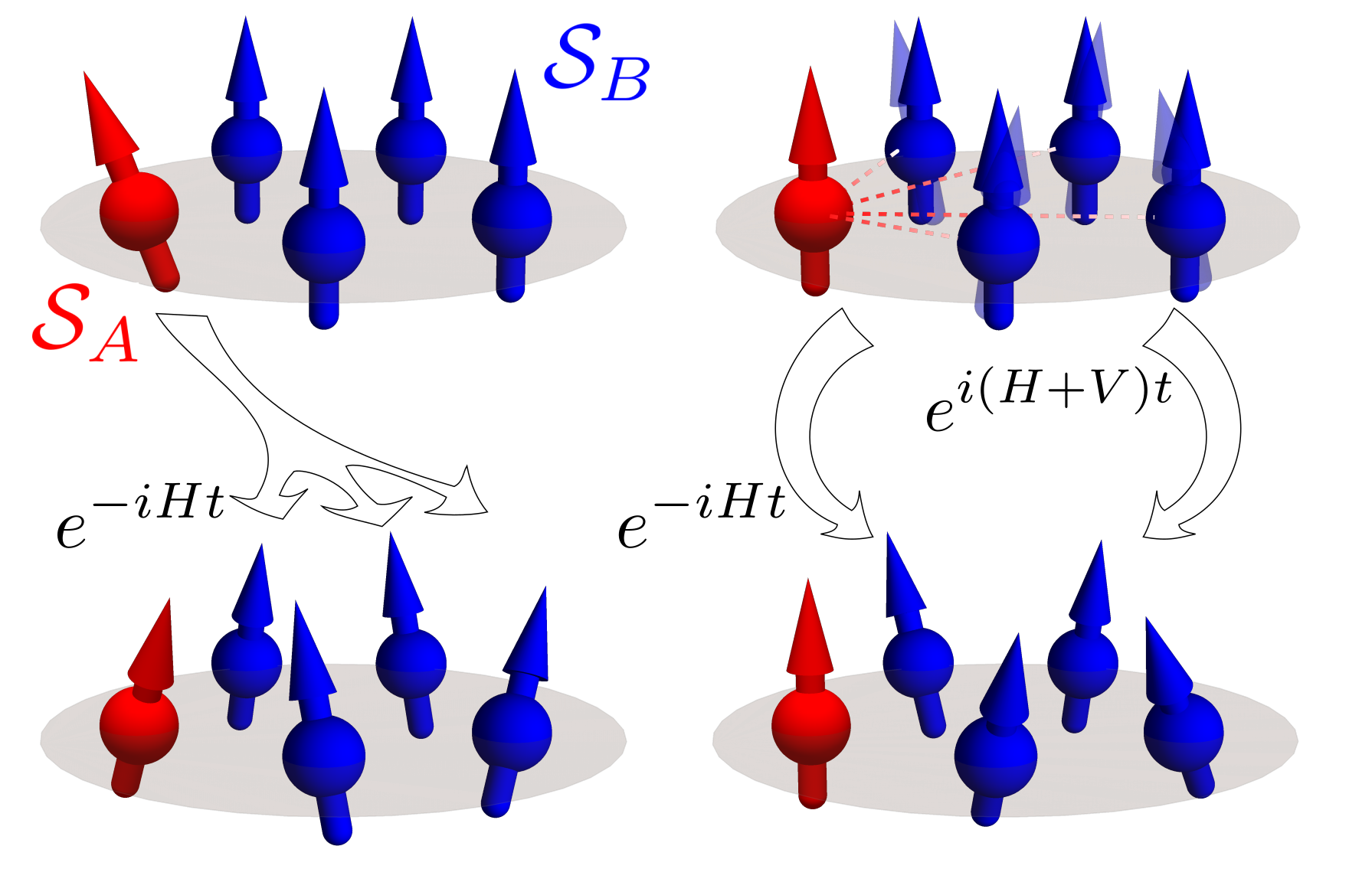}
\caption{\label{fig:system}
An illustration of the system structure under consideration. $\mathcal{S}_A$ is a small local subsystem. $\mathcal{S}_B$ is chosen as the complement of $\mathcal{S}_A$. (Left) A local perturbation on subsystem $\mathcal{S}_A$ spreads over the entire system during time evolution. This can be captured by the decay of OTOC in equation (\ref{eq:otoc}). (Right) The large subsystem $\mathcal{S}_B$ feels an effective random stochastic perturbation $V$ induced by the interaction with the small subsystem $\mathcal{S}_A$. Due to this effective perturbation, the dynamics of $\mathcal{S}_B$ becomes irreversibility. This can be detected by the Loschmidt echo signal in equation (\ref{eq:Loschmidt}). These two processes are shown to be related through the OTOC-LE correspondence in equation (\ref{eq:LE}). Since $\mathcal{S}_A$ is small and local, one can effectively treat $\mathcal{S}_B$ as the total system.}
\end{figure}

We focus on the scenario that $A$ is an operator with support on a small local subsystem $S_A$, while operator $B$ is chosen such that its support $S_B$ is the complement of $S_A$, as illustrated in Fig.~\ref{fig:system}. It is reasonable to expect that choosing operator $B$ in such a manner also captures the spreading of operator $A$ over the entire system, detected by its nonzero projection at later times on the support of operator $B$. Analysis of this particular scenario is also instructive and can be generalized in a similar way to cases where $B$ is a small local operator as well. 

The Haar integral over subsystem operators can be evaluated with the aid of the formula
\begin{equation} \label{eq:Haar}
    \int dA\ A^{\dagger}OA=\frac{1}{d_A}\mathbb{I}_A\otimes\Tr_A(O),
\end{equation}
where $\mathbb{I}_A$ is the identity operator and $\Tr_A$ is the partial trace over subsystem $S_A$. The Haar measure is unique up to a constant multiplicative factor. Here, $d_A$, the dimension of $S_A$ is introduced as a convention to normalize the OTOC \cite{Bhattacharyya2019-nx}.  The proof of the above equation is given in Appendix~\ref{app:Haar}.

The average over all random unitary operators on subsystem $S_A$ gives us
\begin{equation}
\begin{aligned}
       &\int dA\ F_{\beta=0}(t) \equiv \frac{1}{d} \int dA\ \Tr(A^\dagger(t) B^\dagger A(t) B)\\
       = &\frac{1}{d}\frac{1}{d_A} \Tr[ \mathbb{I}_A \otimes \Tr_A(e^{-iHt}B^\dagger e^{iHt})\ e^{-iHt}Be^{iHt}]\\
       = &\frac{1}{d}\frac{1}{d_A} \Tr_B [ \Tr_A(e^{-iHt}B^\dagger e^{iHt})\Tr_A(e^{-iHt}Be^{iHt}) ].
\end{aligned}
\end{equation}
The last line of the above equation involves the reduced dynamics of operator $B$, i.e., $B(-t)\equiv\Tr_A (e^{-iHt}Be^{iHt})$. In order to further perform the average over $B$, we estimate $B(-t)$ in the following way. The total system Hamiltonian, in general, has the structure
\begin{equation}\label{eq:decomp}
\begin{aligned}
       & H=\mathbb{I}_A\otimes H_B + H_A\otimes \mathbb{I}_B + H',\\
    &  H'\equiv\delta\sum_kV_A^k\otimes V_B^k.
\end{aligned}
\end{equation}
Here $V^k_A$'s are Hermitian and orthonormal (with respect to the Hilbert-Schmidt inner product and norm); $V^k_B$'s are Hermitian, orthogonal and have norms on the same order as $H_B$, such that $\delta$ quantifies the relative strength of the coupling. In realistic physical systems, the coupling parameter $\delta\ll 1$. To get the solution of $B(-t)$, we replace the effect of the coupling with an ensemble of random noises $\{V_\alpha\}$ on subsystem $S_B$, namely,
\begin{equation}\label{eq:bdynamics}
\begin{aligned}
      B(-t)&=\Tr_A\left(e^{-iHt}Be^{iHt}\right)\\
         &\propto \overline{e^{-i(H_B+V_\alpha)t}Be^{i(H_B+V_\alpha)t}}.
\end{aligned}
\end{equation}
Here $\alpha$ labels different realizations of the noise and the average taken is over all realizations of the noise operator $V_\alpha$'s. 

The above claim is based on the correspondence between the symptoms of decoherence \cite{Zurek2003-qi,Schlosshauer2007-kp} (process that involves entangling correlations between the system and the environment) and symptoms of the suitable external noises. In Appendix~\ref{app:B}, we back up this claim with a more mathematically rigorous treatment of the noise operators in terms of master equations up to the second order of the coupling parameter $\delta$ \cite{Budini1999-uo}, where it is shown that the noise operators can be chosen as linear combinations of $V_B^k$'s with random coefficients $\pm 1$: 
\begin{equation}\label{eq:noise}
    V_\alpha=\sum_k\pm\delta V_B^k.
\end{equation}

With the aid of the alternative form for the reduced dynamics for $B(-t)$ in Eq.~(\ref{eq:bdynamics}), averaging over operators $B$ can be further performed in the same manner as for the operators $A$. This gives the final expression for the averaged OTOC:

\begin{subequations}
\label{eq:LE}
\begin{equation}\label{eq:LEa}
    \overline{F_{\beta=0}(t)} \approx \overline{ |\langle e^{i(H_B+V_\alpha) t}e^{-i(H_B+V_{\alpha'}) t}\rangle_{\beta=0}|^2}.
\end{equation}
Here, $V_\alpha$ and $V_{\alpha'}$ average over all realizations of the noise operators as given by Eq.~(\ref{eq:noise}). Each term in the average is precisely the Loschmidt echo averaged over a thermal ensemble. 
For systems with large number of degrees of freedom, the ensemble of noises is large and the structures of the noise operators are expected to be not essential. In this case, one can replace the noise average with a single LE to get a coarse-grained version of the above equation, namely,
\begin{equation}\label{eq:LEb}
    \overline{F_{\beta=0}(t)} \approx |\langle e^{i(H_B+V_1) t}e^{-i(H_B+V_2) t}\rangle_{\beta=0}|^2.
\end{equation}
\end{subequations}

Equation~(\ref{eq:LE}) is the main result of this work. Note that: 
(i) As has been mentioned before, the above result generalizes to finite temperatures. We present the full derivation in Appendix~\ref{app:C}, using the same scheme for regularizing the thermal state explored in Refs~\cite{Maldacena2016-mb,Cotler2017-oi,Tsuji2018-td}. This regularization scheme is crucial for the discussion of the bound of the OTOC decay~\cite{Romero-Bermudez2019-lj}. (ii) The perturbations that appear in the LE emerge from the interactions nested in the total Hamiltonian. (iii) The OTOC has several decay regimes, e.g., the early growth $\sim a-\epsilon b  e^{\lambda t} + O(\epsilon^2)$ before the scrambling (Ehrenfest) timescale~\cite{Maldacena2016-mb}, where $a$ and $b$ are order-1 numbers and $\epsilon\ll1$ is a small parameter; or the intermediate pure exponential decay. The only approximation involved in the derivation of Eq.~(\ref{eq:LE}) is the second-order approximation of the coupling parameter $\delta$. It will be shown in the following that the OTOC-LE connection holds in both the scrambling and intermediate decay regimes. Remarkably, the second order of the coupling parameter $\delta^2$ is identified as the small prefactor $\epsilon$ in the
early exponential growth, which determines the scrambling timescale $\sim (1/\lambda)\ln(1/\epsilon)$.

\begin{figure}[t!]
\includegraphics[width=0.95\columnwidth]{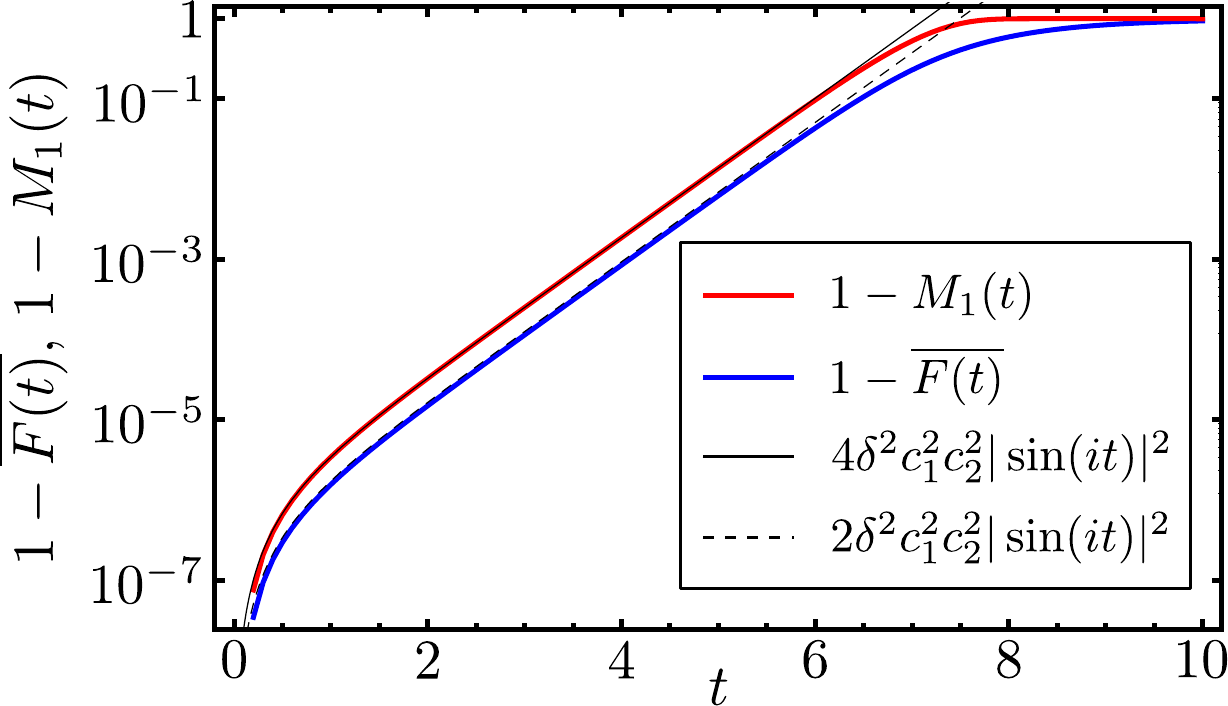}
\caption{\label{fig:IO} Early exponential growth of the OTOC and LE of the IHOs. The average of the OTOC is taken over a pure product state $\psi(x_1)\psi(x_1)\propto e^{-x^2_1/m_2}e^{-x^2_2/m_1}$. The parameters are $m_1=10^5,m_2=1,\omega_1=0,\omega_2=1,\delta=10^{-5}$. The solid blue and red curves correspond to $1-\overline{F(t)}$ and $1-M_1(t)$, respectively. The dashed and solid black curves correspond to $2\delta^2c_1^2c_2^2|\sin(it)|^2$ and $4\delta^2 c_1^2c_2^2|\sin(it)|^2$, respectively, as deduced from Eq.~(\ref{eq:BCH}).}
\end{figure}

\emph{Case I: Early scrambling.---}
To verify the OTOC-LE relation in the scrambling regime, we first study an exactly solvable model consisting of two coupled inverted harmonic oscillators (IHOs), whose Hamiltonian is
\begin{equation}
    \sum_{i=1,2}\left(\frac{1}{2m_i}\hat{p}^2_i-\frac{m_i\omega_i^2}{2}\hat{x}^2_i\right)+\delta \hat{x}_1\hat{x}_2.
\end{equation} 
This model employs IHOs, as they were first used to emulate dynamical instability characteristic of quantum chaos in an exactly solvable system \cite{Blume-Kohout2003-ww}. Their role here is to capture the essential ingredients of the OTOC-LE relation: The OTOC is computed for two (local) operators on the two oscillators, respectively. The parameters are tuned as $m_1\gg m_2$, $\omega_1\ll\omega_2$ and $\delta\ll 1$ such that the dynamics of the first oscillator is much slower than the second one and the coupling is weak. Hence, oscillator $1$ and $2$ mimic the subsystems $S_A$ and $S_B$, respectively.. 

Since the spectrum of the IHOs is not bounded from below, the thermal state is not well defined. We replace the regularized thermal state with a pure state $|\psi_1(x_1)\rangle|\psi_2(x_2)\rangle$. The pure state average of the OTOC has been considered in the literature \cite{Garttner2017-rl}. In this case, the Haar averaged OTOC of the two coupled oscillators can be computed exactly. We depict the result in Fig.~\ref{fig:IO} and delegate the lengthy calculation to Appendix~\ref{app:IHO}, where it is shown that the pure state average carries out in the same manner as a thermal ensemble.

According to Eq.~(\ref{eq:LE}), the OTOC equals the LE of the second oscillator (the large subsystem $S_B$), which admits an exact solution as well. The perturbations emerging from the coupling are $\pm \delta c_1 \hat{x}_2$ with equal probability, where $c^2_1=\langle\psi_1|x^2_1|\psi_1\rangle$. Note that the average of $x^2_1$ appears because of the normalization condition in the decomposition of the interaction in Eq.~(\ref{eq:decomp}). See details about the derivation of the perturbation operators in Appendix~\ref{app:IHO}. As a simple system, the coupled IHO has only one noise operator. Thus the coarse-grained version of the OTOC-LE connection in Eq.~(\ref{eq:LEb}) is not reliable. We instead explicitly use the exact form in Eq.~(\ref{eq:LEa}), the right-hand side of which reduces to
\begin{equation}
   M(t)=\frac{1}{2}+\frac{1}{2}|\langle\psi_2|e^{i(H_2+\delta c_1 \hat{x}_2)t}e^{-i(H_2-\delta c_1 \hat{x}_2)t}|\psi_2\rangle|^2.
\end{equation}
Denote $M_1(t)\equiv |\langle\psi|e^{i(H_2+\delta c_1 \hat{x}_2)t}e^{-i(H_2-\delta c_1 \hat{x}_2)t}|\psi\rangle|^2$. 
To extract the early exponential growth, consider the quantity $1-\overline{F(t)}\approx 1-M(t)= \frac{1}{2}\left(1-M_1(t)\right)$. This implies that, in the early growth regime, the averaged OTOC $\overline{F(t)}$ has the same exponential growth rate as $M_1(t)$, but a prefactor half of the latter. Note that the growth of the LE $M(t)$ does not saturate to one. The reason is that the OTOC-LE connection is exact up to second order of the coupling parameter $\delta$. As will be demonstrated in the following, $\delta^2$ plays the role of the prefactor $\epsilon$ of the OTOC early growth $\sim \epsilon e^{\lambda t}$, which is precisely reflected in the early growth of the LE. The saturation of the OTOC is induced by higher orders of $\epsilon$. 

To see the significance of the coupling strength, we expand $M_1(t)$ to second order of $\delta$ using the Baker-Campbell-Hausdorff (BCH) formula, which gives
\begin{equation}\label{eq:BCH}
    M_1(t)=1-\delta^2\frac{ 4c_1^2c_2^2}{(i\omega_2)^2}\sin^2{(i\omega_2t)},
\end{equation}
where $c^2_2=\langle\psi_2|x_2^2|\psi_2\rangle$. This describes an exponential growth with rate $2\omega_2$. The second order of the coupling parameter plays the role of the prefactor in the early exponential Lyapunov growth. Derivations for the exact solution of $M_1(t)$ and its second-order BCH expansion are presented in Appendix~\ref{app:IHO}. Figure~\ref{fig:IO} depicts the exact solutions of the Haar averaged OTOC, $M_1(t)$ as well as the exponential growth extracted from the second-order expansion in Eq.~(\ref{eq:BCH}). The OTOC-LE connection and the early scrambling are clearly revealed. 

\begin{figure}[t!]
  \centering
  \includegraphics[width=\columnwidth]{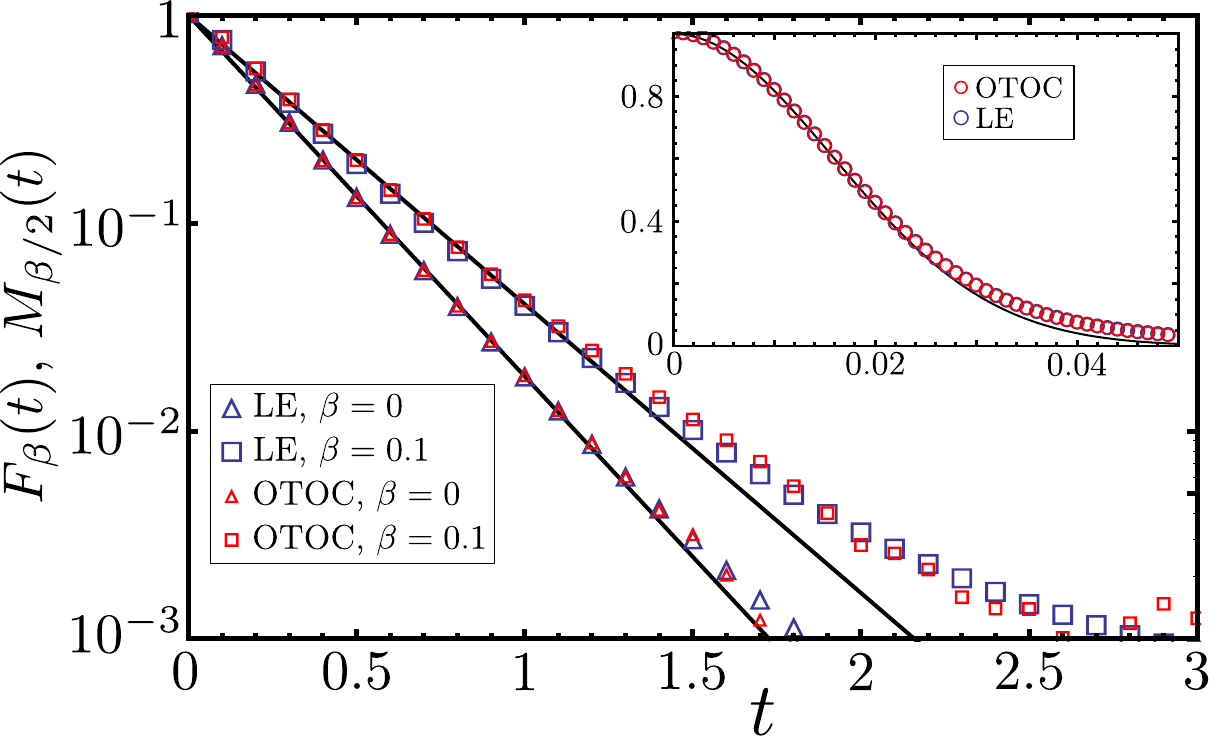}
  \caption{Numerical simulations of the OTOC and LE for the random matrix model. The Hilbert space dimensions of subsystem $S_A$ and $S_B$ are $2$ and $2^9$, respectively. Red and blue marks correspond to data of the OTOC and LE, respectively. In the main figure, the coupling parameter is fixed at $\delta=0.1$, for which the decay is exponential. Triangles and squares correspond to $\beta=0$ and $0.1$, respectively. Inset: Gaussian decay of the OTOC and LE at coupling strength $\delta=0.5$. The temperature is fixed at infinity. Black lines are best fits to exponential or Gaussian curves.}\label{fig:RM}
\end{figure}

\emph{Case II: Intermediate decay.---} The early exponential growth of the OTOC has been predicted and observed in various platforms (See discussions in Ref.~\cite{Xu2020-xo} and the references therein). However, in a variety of systems, especially finite-size lattice systems, the scrambling time is typically too small to be reliably visible. Instead, the decay of the OTOC is a pure exponential. To reveal the OTOC-LE relation in such an intermediate decay regime, we study a random matrix model. The model Hamiltonian takes the general form Eq.~(\ref{eq:decomp}). The noninteracting part of the Hamiltonian are random matrices from the standard Gaussian unitary ensemble (GUE), whose matrix elements have independent real and imaginary parts as Gaussian random numbers with zero mean and unit variance. The small subsystem $S_A$ is chosen as a single qubit. The coupling matrix $V^k_A$ in the decomposition Eq.~(\ref{eq:decomp}) are chosen as Pauli matrices, i.e., $\{I,\sigma_x,\sigma_y,\sigma_z\}$, while $V^k_B$ on the large subsystem $S_B$ are drawn from the GUE.

The random matrix model has enough complexity to make the OTOC insensitive to the choice of operators. This allows us to numerically simulate the evolution of the OTOC for two random Hermitian operators on the corresponding subsystems. The LE can be computed using the coarse-grained version of the OTOC-LE relation in Eq.~(\ref{eq:LEb}). The effective perturbation operators, by Eq.~(\ref{eq:noise}), are the sum of four random matrices from the GUE and, therefore, have matrix elements with variance $4\delta^2$. 

It is well known that in the intermediate regime the decay of the LE depends on the strength of the perturbation; i.e., for small perturbations the decay is exponential, while for large perturbations the decay can be Gaussian. Since the relative coupling strength decreases with the system size, exponential decays are typically expected in the thermodynamics limit. However, for small size systems Gaussian decays might be observed. This explains the Gaussian decay of the OTOC in finite-size systems, e.g., the Sachdev-Ye-Kitaev (SYK) model \cite{Sachdev1993-ka,Kitaev2015,Maldacena2016-yh}, for which in the large-$N$ limit the OTOC switches from an early growth to an intermediate exponential decay. As another application of our theory, we present a detailed discussion of the SYK model in Appendix~\ref{app:SYK}.

The numerical simulations for the random matrix model are presented in Fig.~\ref{fig:RM}, where the OTOC and LE are shown to match very well; both exponential and Gaussian decays are revealed.

\emph{The bound on chaos.---} 
In the scrambling regime, the early exponential growth rate was conjectured to be universally bounded by the temperature, $\lambda\le2\pi/\beta$. This conjecture has been proven under a strong physical assumption that the time-ordered correlators factorize. As a consequence of that assumption, the magnitude of the normalized OTOC is always bounded by unity in the analytic regime on the complex time domain (see Refs.~\cite{Maldacena2016-mb,Tsuji2018-td} for detailed discussions). As another remarkable application of the OTOC-LE connection, we note that the factorization assumption can be removed for the Haar averaged OTOC, which is attributed to the fact that, in the corresponding complex time domain, the LE appears to be averaged over a thermal state with a positive effective temperature. We refer the interested reader to Appendix~\ref{app:C} for a detailed discussion.

\emph{Summary.---} We have demonstrated the connection between two distinct areas of the dynamical quantum chaos, namely, the emerging field of the out-of-time-order correlators and the relatively more developed field of the Loschmidt echo. This relation not only allows a more general understanding of the universal properties of the OTOC but also provides new insights into both subjects. Two models were used to reveal the OTOC-LE relation in the scrambling regime and the intermediate exponential decay regime. Implications to the bound on the decay rate were also discussed. Future research could generalize the connection to higher-dimensional systems with spatial structures.

\begin{acknowledgements}
This research was supported by the U.S. Department of Energy, Office of Science, Basic Energy Sciences, Materials Sciences and Engineering Division, Condensed Matter Theory Program. WHZ and LC were also supported in part by the LDRD program at the Los Alamos National Laboratory. WHZ acknowledges partial support by the Foundational Questions Institute grant FQXi-1821, and Franklin Fetzer Fund, a donor advised fund of the Silicon Valley Community Foundation. LC acknowledges support by the DOE through the J. Robert Oppenheimer fellowship. BY and WHZ thank Adolfo del Campo and Nikolai Sinitsyn for helpful discussions.
\end{acknowledgements}

\bibliography{reference}

\begin{thebibliography}{62}%
\makeatletter
\providecommand \@ifxundefined [1]{%
 \@ifx{#1\undefined}
}%
\providecommand \@ifnum [1]{%
 \ifnum #1\expandafter \@firstoftwo
 \else \expandafter \@secondoftwo
 \fi
}%
\providecommand \@ifx [1]{%
 \ifx #1\expandafter \@firstoftwo
 \else \expandafter \@secondoftwo
 \fi
}%
\providecommand \natexlab [1]{#1}%
\providecommand \enquote  [1]{``#1''}%
\providecommand \bibnamefont  [1]{#1}%
\providecommand \bibfnamefont [1]{#1}%
\providecommand \citenamefont [1]{#1}%
\providecommand \href@noop [0]{\@secondoftwo}%
\providecommand \href [0]{\begingroup \@sanitize@url \@href}%
\providecommand \@href[1]{\@@startlink{#1}\@@href}%
\providecommand \@@href[1]{\endgroup#1\@@endlink}%
\providecommand \@sanitize@url [0]{\catcode `\\12\catcode `\$12\catcode
  `\&12\catcode `\#12\catcode `\^12\catcode `\_12\catcode `\%12\relax}%
\providecommand \@@startlink[1]{}%
\providecommand \@@endlink[0]{}%
\providecommand \url  [0]{\begingroup\@sanitize@url \@url }%
\providecommand \@url [1]{\endgroup\@href {#1}{\urlprefix }}%
\providecommand \urlprefix  [0]{URL }%
\providecommand \Eprint [0]{\href }%
\providecommand \doibase [0]{http://dx.doi.org/}%
\providecommand \selectlanguage [0]{\@gobble}%
\providecommand \bibinfo  [0]{\@secondoftwo}%
\providecommand \bibfield  [0]{\@secondoftwo}%
\providecommand \translation [1]{[#1]}%
\providecommand \BibitemOpen [0]{}%
\providecommand \bibitemStop [0]{}%
\providecommand \bibitemNoStop [0]{.\EOS\space}%
\providecommand \EOS [0]{\spacefactor3000\relax}%
\providecommand \BibitemShut  [1]{\csname bibitem#1\endcsname}%
\let\auto@bib@innerbib\@empty
\bibitem [{\citenamefont {Haake}(2006)}]{Haake2006-ic}%
  \BibitemOpen
  \bibfield  {author} {\bibinfo {author} {\bibfnamefont {F.}~\bibnamefont
  {Haake}},\ }\href@noop {} {\emph {\bibinfo {title} {Quantum Signatures of
  Chaos}}}\ (\bibinfo  {publisher} {Springer-Verlag},\ \bibinfo {address}
  {Berlin, Heidelberg},\ \bibinfo {year} {2006})\BibitemShut {NoStop}%
\bibitem [{\citenamefont {Berry}(1989)}]{Berry2006-gh}%
  \BibitemOpen
  \bibfield  {author} {\bibinfo {author} {\bibfnamefont {M.}~\bibnamefont
  {Berry}},\ }\bibfield  {title} {\enquote {\bibinfo {title} {Quantum chaology,
  not quantum chaos},}\ }\href {\doibase 10.1088/0031-8949/40/3/013} {\bibfield
   {journal} {\bibinfo  {journal} {Phys. Scr.}\ }\textbf {\bibinfo {volume}
  {40}},\ \bibinfo {pages} {335} (\bibinfo {year} {1989})}\BibitemShut
  {NoStop}%
\bibitem [{\citenamefont {Peres}(2002)}]{Peres2002-ew}%
  \BibitemOpen
  \bibinfo {editor} {\bibfnamefont {A.}~\bibnamefont {Peres}},\ ed.,\
  \href@noop {} {\emph {\bibinfo {title} {Quantum Theory: Concepts and
  Methods}}},\ Fundamental Theories of Physics\ (\bibinfo  {publisher}
  {Springer},\ \bibinfo {year} {2002})\BibitemShut {NoStop}%
\bibitem [{\citenamefont {Zurek}\ and\ \citenamefont
  {Paz}(1994)}]{Zurek1994-om}%
  \BibitemOpen
  \bibfield  {author} {\bibinfo {author} {\bibfnamefont {W.~H.}\ \bibnamefont
  {Zurek}}\ and\ \bibinfo {author} {\bibfnamefont {J.~P.}\ \bibnamefont
  {Paz}},\ }\bibfield  {title} {\enquote {\bibinfo {title} {Decoherence, chaos,
  and the second law},}\ }\href {\doibase 10.1103/PhysRevLett.72.2508}
  {\bibfield  {journal} {\bibinfo  {journal} {Phys. Rev. Lett.}\ }\textbf
  {\bibinfo {volume} {72}},\ \bibinfo {pages} {2508--2511} (\bibinfo {year}
  {1994})}\BibitemShut {NoStop}%
\bibitem [{\citenamefont {Zurek}(1998)}]{Zurek2006-vj}%
  \BibitemOpen
  \bibfield  {author} {\bibinfo {author} {\bibfnamefont {W.~H.}\ \bibnamefont
  {Zurek}},\ }\bibfield  {title} {\enquote {\bibinfo {title} {{Decoherence,
  Chaos, Quantum-Classical Correspondence, and the Algorithmic Arrow of
  Time}},}\ }\href {\doibase 10.1238/Physica.Topical.076a00186} {\bibfield
  {journal} {\bibinfo  {journal} {Physica Scripta Volume T}\ }\textbf {\bibinfo
  {volume} {76}},\ \bibinfo {pages} {186--198} (\bibinfo {year}
  {1998})}\BibitemShut {NoStop}%
\bibitem [{\citenamefont {Levstein}\ \emph {et~al.}(1998)\citenamefont
  {Levstein}, \citenamefont {Usaj},\ and\ \citenamefont
  {Pastawski}}]{Levstein1998-um}%
  \BibitemOpen
  \bibfield  {author} {\bibinfo {author} {\bibfnamefont {P.~R.}\ \bibnamefont
  {Levstein}}, \bibinfo {author} {\bibfnamefont {G.}~\bibnamefont {Usaj}}, \
  and\ \bibinfo {author} {\bibfnamefont {H.~M.}\ \bibnamefont {Pastawski}},\
  }\bibfield  {title} {\enquote {\bibinfo {title} {Attenuation of polarization
  echoes in nuclear magnetic resonance: A study of the emergence of dynamical
  irreversibility in many-body quantum systems},}\ }\href {\doibase
  10.1063/1.475664} {\bibfield  {journal} {\bibinfo  {journal} {J. Chem.
  Phys.}\ }\textbf {\bibinfo {volume} {108}},\ \bibinfo {pages} {2718--2724}
  (\bibinfo {year} {1998})}\BibitemShut {NoStop}%
\bibitem [{\citenamefont {Jalabert}\ and\ \citenamefont
  {Pastawski}(2001)}]{Jalabert2001-kn}%
  \BibitemOpen
  \bibfield  {author} {\bibinfo {author} {\bibfnamefont {R.~A.}\ \bibnamefont
  {Jalabert}}\ and\ \bibinfo {author} {\bibfnamefont {H.~M.}\ \bibnamefont
  {Pastawski}},\ }\bibfield  {title} {\enquote {\bibinfo {title}
  {Environment-independent decoherence rate in classically chaotic systems},}\
  }\href {\doibase 10.1103/PhysRevLett.86.2490} {\bibfield  {journal} {\bibinfo
   {journal} {Phys. Rev. Lett.}\ }\textbf {\bibinfo {volume} {86}},\ \bibinfo
  {pages} {2490--2493} (\bibinfo {year} {2001})}\BibitemShut {NoStop}%
\bibitem [{\citenamefont {Cucchietti}\ \emph {et~al.}(2003)\citenamefont
  {Cucchietti}, \citenamefont {Dalvit}, \citenamefont {Paz},\ and\
  \citenamefont {Zurek}}]{Cucchietti2003-hh}%
  \BibitemOpen
  \bibfield  {author} {\bibinfo {author} {\bibfnamefont {F.~M.}\ \bibnamefont
  {Cucchietti}}, \bibinfo {author} {\bibfnamefont {D.~A.~R.}\ \bibnamefont
  {Dalvit}}, \bibinfo {author} {\bibfnamefont {J.~P.}\ \bibnamefont {Paz}}, \
  and\ \bibinfo {author} {\bibfnamefont {W.~H.}\ \bibnamefont {Zurek}},\
  }\bibfield  {title} {\enquote {\bibinfo {title} {{Decoherence and the
  Loschmidt echo}},}\ }\href {\doibase 10.1103/PhysRevLett.91.210403}
  {\bibfield  {journal} {\bibinfo  {journal} {Phys. Rev. Lett.}\ }\textbf
  {\bibinfo {volume} {91}},\ \bibinfo {pages} {210403} (\bibinfo {year}
  {2003})}\BibitemShut {NoStop}%
\bibitem [{\citenamefont {Kohler}\ and\ \citenamefont
  {Recher}(2012)}]{Kohler2012-oz}%
  \BibitemOpen
  \bibfield  {author} {\bibinfo {author} {\bibfnamefont {H.}~\bibnamefont
  {Kohler}}\ and\ \bibinfo {author} {\bibfnamefont {C.}~\bibnamefont
  {Recher}},\ }\bibfield  {title} {\enquote {\bibinfo {title} {Fidelity and
  level correlations in the transition from regularity to chaos},}\ }\href
  {\doibase 10.1209/0295-5075/98/10005} {\bibfield  {journal} {\bibinfo
  {journal} {EPL}\ }\textbf {\bibinfo {volume} {98}},\ \bibinfo {pages} {10005}
  (\bibinfo {year} {2012})}\BibitemShut {NoStop}%
\bibitem [{\citenamefont {{S{\'a}nchez}}\ \emph {et~al.}(2016)\citenamefont
  {{S{\'a}nchez}}, \citenamefont {{Levstein}}, \citenamefont {{Buljubasich}},
  \citenamefont {{Pastawski}},\ and\ \citenamefont
  {{Chattah}}}]{Sanchez2016-mx}%
  \BibitemOpen
  \bibfield  {author} {\bibinfo {author} {\bibfnamefont {C.~M.}\ \bibnamefont
  {{S{\'a}nchez}}}, \bibinfo {author} {\bibfnamefont {P.~R.}\ \bibnamefont
  {{Levstein}}}, \bibinfo {author} {\bibfnamefont {L.}~\bibnamefont
  {{Buljubasich}}}, \bibinfo {author} {\bibfnamefont {H.~M.}\ \bibnamefont
  {{Pastawski}}}, \ and\ \bibinfo {author} {\bibfnamefont {A.~K.}\ \bibnamefont
  {{Chattah}}},\ }\bibfield  {title} {\enquote {\bibinfo {title} {{Quantum
  dynamics of excitations and decoherence in many-spin systems detected with
  Loschmidt echoes: its relation to their spreading through the Hilbert
  space}},}\ }\href {\doibase 10.1098/rsta.2015.0155} {\bibfield  {journal}
  {\bibinfo  {journal} {Philosophical Transactions of the Royal Society of
  London Series A}\ }\textbf {\bibinfo {volume} {374}},\ \bibinfo {eid}
  {20150155} (\bibinfo {year} {2016})}\BibitemShut {NoStop}%
\bibitem [{\citenamefont {Chenu}\ \emph {et~al.}(2018)\citenamefont {Chenu},
  \citenamefont {Egusquiza}, \citenamefont {Molina-Vilaplana},\ and\
  \citenamefont {del Campo}}]{Chenu2018-sm}%
  \BibitemOpen
  \bibfield  {author} {\bibinfo {author} {\bibfnamefont {A.}~\bibnamefont
  {Chenu}}, \bibinfo {author} {\bibfnamefont {I.~L.}\ \bibnamefont
  {Egusquiza}}, \bibinfo {author} {\bibfnamefont {J}~\bibnamefont
  {Molina-Vilaplana}}, \ and\ \bibinfo {author} {\bibfnamefont
  {A.}~\bibnamefont {del Campo}},\ }\bibfield  {title} {\enquote {\bibinfo
  {title} {Quantum work statistics, loschmidt echo and information
  scrambling},}\ }\href {\doibase 10.1038/s41598-018-30982-w} {\bibfield
  {journal} {\bibinfo  {journal} {Sci. Rep.}\ }\textbf {\bibinfo {volume}
  {8}},\ \bibinfo {pages} {12634} (\bibinfo {year} {2018})}\BibitemShut
  {NoStop}%
\bibitem [{\citenamefont {Chenu}\ \emph {et~al.}(2019)\citenamefont {Chenu},
  \citenamefont {Molina-Vilaplana},\ and\ \citenamefont {del
  Campo}}]{Chenu2019-ue}%
  \BibitemOpen
  \bibfield  {author} {\bibinfo {author} {\bibfnamefont {A}~\bibnamefont
  {Chenu}}, \bibinfo {author} {\bibfnamefont {J.}~\bibnamefont
  {Molina-Vilaplana}}, \ and\ \bibinfo {author} {\bibfnamefont
  {A.}~\bibnamefont {del Campo}},\ }\bibfield  {title} {\enquote {\bibinfo
  {title} {Work statistics, loschmidt echo and information scrambling in
  chaotic quantum systems},}\ }\href {\doibase 10.22331/q-2019-03-04-127}
  {\bibfield  {journal} {\bibinfo  {journal} {Quantum}\ }\textbf {\bibinfo
  {volume} {3}},\ \bibinfo {pages} {127} (\bibinfo {year} {2019})}\BibitemShut
  {NoStop}%
\bibitem [{\citenamefont {S{\'a}nchez}\ \emph {et~al.}(2020)\citenamefont
  {S{\'a}nchez}, \citenamefont {Chattah}, \citenamefont {Wei}, \citenamefont
  {Buljubasich}, \citenamefont {Cappellaro},\ and\ \citenamefont
  {Pastawski}}]{Sanchez2019-rs}%
  \BibitemOpen
  \bibfield  {author} {\bibinfo {author} {\bibfnamefont {C~M}\ \bibnamefont
  {S{\'a}nchez}}, \bibinfo {author} {\bibfnamefont {A~K}\ \bibnamefont
  {Chattah}}, \bibinfo {author} {\bibfnamefont {K~X}\ \bibnamefont {Wei}},
  \bibinfo {author} {\bibfnamefont {L}~\bibnamefont {Buljubasich}}, \bibinfo
  {author} {\bibfnamefont {P}~\bibnamefont {Cappellaro}}, \ and\ \bibinfo
  {author} {\bibfnamefont {H~M}\ \bibnamefont {Pastawski}},\ }\bibfield
  {title} {\enquote {\bibinfo {title} {Perturbation independent decay of the
  loschmidt echo in a {Many-Body} system},}\ }\href {\doibase
  10.1103/PhysRevLett.124.030601} {\bibfield  {journal} {\bibinfo  {journal}
  {Phys. Rev. Lett.}\ }\textbf {\bibinfo {volume} {124}},\ \bibinfo {pages}
  {030601} (\bibinfo {year} {2020})}\BibitemShut {NoStop}%
\bibitem [{\citenamefont {Goussev}\ \emph {et~al.}(2012)\citenamefont
  {Goussev}, \citenamefont {Jalabert}, \citenamefont {Pastawski},\ and\
  \citenamefont {Wisniacki}}]{Goussev2012-ep}%
  \BibitemOpen
  \bibfield  {author} {\bibinfo {author} {\bibfnamefont {A.}~\bibnamefont
  {Goussev}}, \bibinfo {author} {\bibfnamefont {R.~A.}\ \bibnamefont
  {Jalabert}}, \bibinfo {author} {\bibfnamefont {H.~M.}\ \bibnamefont
  {Pastawski}}, \ and\ \bibinfo {author} {\bibfnamefont {D.}~\bibnamefont
  {Wisniacki}},\ }\enquote {\bibinfo {title} {Loschmidt echo},}\ in\ \href@noop
  {} {\emph {\bibinfo {booktitle} {Scholarpedia}}}\ (\bibinfo  {publisher}
  {Scholarpedia},\ \bibinfo {year} {2012})\ p.\ \bibinfo {pages}
  {7(8):11687}\BibitemShut {NoStop}%
\bibitem [{\citenamefont {Gorin}\ \emph {et~al.}(2006)\citenamefont {Gorin},
  \citenamefont {Prosen}, \citenamefont {Seligman},\ and\ \citenamefont {{\v
  Z}nidari{\v c}}}]{Gorin2006-ro}%
  \BibitemOpen
  \bibfield  {author} {\bibinfo {author} {\bibfnamefont {T.}~\bibnamefont
  {Gorin}}, \bibinfo {author} {\bibfnamefont {T.}~\bibnamefont {Prosen}},
  \bibinfo {author} {\bibfnamefont {T.~H.}\ \bibnamefont {Seligman}}, \ and\
  \bibinfo {author} {\bibfnamefont {M.}~\bibnamefont {{\v Z}nidari{\v c}}},\
  }\bibfield  {title} {\enquote {\bibinfo {title} {{Dynamics of Loschmidt
  echoes and fidelity decay}},}\ }\href {\doibase
  10.1016/j.physrep.2006.09.003} {\bibfield  {journal} {\bibinfo  {journal}
  {Phys. Rep.}\ }\textbf {\bibinfo {volume} {435}},\ \bibinfo {pages} {33--156}
  (\bibinfo {year} {2006})}\BibitemShut {NoStop}%
\bibitem [{\citenamefont {Larkin}\ and\ \citenamefont
  {Ovchinnikov}(1969)}]{Larkin1969}%
  \BibitemOpen
  \bibfield  {author} {\bibinfo {author} {\bibfnamefont {A.}~\bibnamefont
  {Larkin}}\ and\ \bibinfo {author} {\bibfnamefont {Y.~N.}\ \bibnamefont
  {Ovchinnikov}},\ }\bibfield  {title} {\enquote {\bibinfo {title}
  {Quasiclassical method in the theory of superconductivity},}\ }\href@noop {}
  {\bibfield  {journal} {\bibinfo  {journal} {Sov. Phys. JETP}\ }\textbf
  {\bibinfo {volume} {28}},\ \bibinfo {pages} {1200} (\bibinfo {year}
  {1969})}\BibitemShut {NoStop}%
\bibitem [{\citenamefont {Kitaev}()}]{Kitaev2015}%
  \BibitemOpen
  \bibfield  {author} {\bibinfo {author} {\bibfnamefont {A.}~\bibnamefont
  {Kitaev}},\ }\href@noop {} {\bibinfo  {journal} {A simple model of quantum
  holography, in Proceedings of the KITP Program: Entanglement in
  Strongly-Correlated Quantum Matter, 2015 (Kavli Institute for Theoretical
  Physics, Santa Barbara, 2015), Vol. 7}\ }\BibitemShut {NoStop}%
\bibitem [{\citenamefont {Shenker}\ and\ \citenamefont
  {Stanford}(2014{\natexlab{a}})}]{Shenker2014-lz}%
  \BibitemOpen
\bibfield  {journal} {  }\bibfield  {author} {\bibinfo {author} {\bibfnamefont
  {S.~H.}\ \bibnamefont {Shenker}}\ and\ \bibinfo {author} {\bibfnamefont
  {D.}~\bibnamefont {Stanford}},\ }\bibfield  {title} {\enquote {\bibinfo
  {title} {Black holes and the butterfly effect},}\ }\href {\doibase
  10.1007/JHEP03(2014)067} {\bibfield  {journal} {\bibinfo  {journal} {J. High
  Energy Phys.}\ }\textbf {\bibinfo {volume} {03}},\ \bibinfo {pages} {067}
  (\bibinfo {year} {2014}{\natexlab{a}})}\BibitemShut {NoStop}%
\bibitem [{\citenamefont {Shenker}\ and\ \citenamefont
  {Stanford}(2014{\natexlab{b}})}]{Shenker2014-ep}%
  \BibitemOpen
  \bibfield  {author} {\bibinfo {author} {\bibfnamefont {S.~H.}\ \bibnamefont
  {Shenker}}\ and\ \bibinfo {author} {\bibfnamefont {D.}~\bibnamefont
  {Stanford}},\ }\bibfield  {title} {\enquote {\bibinfo {title} {Multiple
  shocks},}\ }\href {\doibase 10.1007/JHEP12(2014)046} {\bibfield  {journal}
  {\bibinfo  {journal} {J. High Energy Phys.}\ }\textbf {\bibinfo {volume}
  {12}},\ \bibinfo {pages} {046} (\bibinfo {year}
  {2014}{\natexlab{b}})}\BibitemShut {NoStop}%
\bibitem [{\citenamefont {Roberts}\ \emph {et~al.}(2015)\citenamefont
  {Roberts}, \citenamefont {Stanford},\ and\ \citenamefont
  {Susskind}}]{Roberts2015-bo}%
  \BibitemOpen
  \bibfield  {author} {\bibinfo {author} {\bibfnamefont {D.~A.}\ \bibnamefont
  {Roberts}}, \bibinfo {author} {\bibfnamefont {D.}~\bibnamefont {Stanford}}, \
  and\ \bibinfo {author} {\bibfnamefont {L.}~\bibnamefont {Susskind}},\
  }\bibfield  {title} {\enquote {\bibinfo {title} {Localized shocks},}\ }\href
  {\doibase 10.1007/JHEP03(2015)051} {\bibfield  {journal} {\bibinfo  {journal}
  {J. High Energy Phys.}\ }\textbf {\bibinfo {volume} {03}},\ \bibinfo {pages}
  {051} (\bibinfo {year} {2015})}\BibitemShut {NoStop}%
\bibitem [{\citenamefont {Roberts}\ and\ \citenamefont
  {Stanford}(2015)}]{Roberts2015-xu}%
  \BibitemOpen
  \bibfield  {author} {\bibinfo {author} {\bibfnamefont {D.~A.}\ \bibnamefont
  {Roberts}}\ and\ \bibinfo {author} {\bibfnamefont {D.}~\bibnamefont
  {Stanford}},\ }\bibfield  {title} {\enquote {\bibinfo {title} {Diagnosing
  chaos using {Four-Point} functions in {Two-Dimensional} conformal field
  theory},}\ }\href {\doibase 10.1103/PhysRevLett.115.131603} {\bibfield
  {journal} {\bibinfo  {journal} {Phys. Rev. Lett.}\ }\textbf {\bibinfo
  {volume} {115}},\ \bibinfo {pages} {131603} (\bibinfo {year}
  {2015})}\BibitemShut {NoStop}%
\bibitem [{\citenamefont {Blake}(2016)}]{Blake2016-eg}%
  \BibitemOpen
  \bibfield  {author} {\bibinfo {author} {\bibfnamefont {M.}~\bibnamefont
  {Blake}},\ }\bibfield  {title} {\enquote {\bibinfo {title} {Universal charge
  diffusion and the butterfly effect in holographic theories},}\ }\href
  {\doibase 10.1103/PhysRevLett.117.091601} {\bibfield  {journal} {\bibinfo
  {journal} {Phys. Rev. Lett.}\ }\textbf {\bibinfo {volume} {117}},\ \bibinfo
  {pages} {091601} (\bibinfo {year} {2016})}\BibitemShut {NoStop}%
\bibitem [{\citenamefont {Roberts}\ and\ \citenamefont
  {Swingle}(2016)}]{Roberts2016-nd}%
  \BibitemOpen
  \bibfield  {author} {\bibinfo {author} {\bibfnamefont {D.~A.}\ \bibnamefont
  {Roberts}}\ and\ \bibinfo {author} {\bibfnamefont {B.}~\bibnamefont
  {Swingle}},\ }\bibfield  {title} {\enquote {\bibinfo {title} {{Lieb-Robinson}
  bound and the butterfly effect in quantum field theories},}\ }\href {\doibase
  10.1103/PhysRevLett.117.091602} {\bibfield  {journal} {\bibinfo  {journal}
  {Phys. Rev. Lett.}\ }\textbf {\bibinfo {volume} {117}},\ \bibinfo {pages}
  {091602} (\bibinfo {year} {2016})}\BibitemShut {NoStop}%
\bibitem [{\citenamefont {Maldacena}\ \emph {et~al.}(2016)\citenamefont
  {Maldacena}, \citenamefont {Shenker},\ and\ \citenamefont
  {Stanford}}]{Maldacena2016-mb}%
  \BibitemOpen
  \bibfield  {author} {\bibinfo {author} {\bibfnamefont {J.}~\bibnamefont
  {Maldacena}}, \bibinfo {author} {\bibfnamefont {S.~H.}\ \bibnamefont
  {Shenker}}, \ and\ \bibinfo {author} {\bibfnamefont {D.}~\bibnamefont
  {Stanford}},\ }\bibfield  {title} {\enquote {\bibinfo {title} {A bound on
  chaos},}\ }\href {\doibase 10.1007/JHEP08(2016)106} {\bibfield  {journal}
  {\bibinfo  {journal} {J. High Energy Phys.}\ }\textbf {\bibinfo {volume}
  {08}},\ \bibinfo {pages} {106} (\bibinfo {year} {2016})}\BibitemShut
  {NoStop}%
\bibitem [{\citenamefont {Fu}\ and\ \citenamefont {Sachdev}(2016)}]{Fu2016-dg}%
  \BibitemOpen
  \bibfield  {author} {\bibinfo {author} {\bibfnamefont {W.}~\bibnamefont
  {Fu}}\ and\ \bibinfo {author} {\bibfnamefont {S.}~\bibnamefont {Sachdev}},\
  }\bibfield  {title} {\enquote {\bibinfo {title} {Numerical study of fermion
  and boson models with infinite-range random interactions},}\ }\href {\doibase
  10.1103/PhysRevB.94.035135} {\bibfield  {journal} {\bibinfo  {journal} {Phys.
  Rev. B}\ }\textbf {\bibinfo {volume} {94}},\ \bibinfo {pages} {035135}
  (\bibinfo {year} {2016})}\BibitemShut {NoStop}%
\bibitem [{\citenamefont {Hosur}\ \emph {et~al.}(2016)\citenamefont {Hosur},
  \citenamefont {Qi}, \citenamefont {Roberts},\ and\ \citenamefont
  {Yoshida}}]{Hosur2016-gb}%
  \BibitemOpen
  \bibfield  {author} {\bibinfo {author} {\bibfnamefont {P.}~\bibnamefont
  {Hosur}}, \bibinfo {author} {\bibfnamefont {X.-L.}\ \bibnamefont {Qi}},
  \bibinfo {author} {\bibfnamefont {D.~A.}\ \bibnamefont {Roberts}}, \ and\
  \bibinfo {author} {\bibfnamefont {B.}~\bibnamefont {Yoshida}},\ }\bibfield
  {title} {\enquote {\bibinfo {title} {Chaos in quantum channels},}\ }\href
  {\doibase 10.1007/JHEP02(2016)004} {\bibfield  {journal} {\bibinfo  {journal}
  {J. High Energy Phys.}\ }\textbf {\bibinfo {volume} {02}},\ \bibinfo {pages}
  {0044} (\bibinfo {year} {2016})}\BibitemShut {NoStop}%
\bibitem [{\citenamefont {Tsuji}\ \emph {et~al.}(2018)\citenamefont {Tsuji},
  \citenamefont {Shitara},\ and\ \citenamefont {Ueda}}]{Tsuji2018-td}%
  \BibitemOpen
  \bibfield  {author} {\bibinfo {author} {\bibfnamefont {N.}~\bibnamefont
  {Tsuji}}, \bibinfo {author} {\bibfnamefont {T.}~\bibnamefont {Shitara}}, \
  and\ \bibinfo {author} {\bibfnamefont {M.}~\bibnamefont {Ueda}},\ }\bibfield
  {title} {\enquote {\bibinfo {title} {Bound on the exponential growth rate of
  out-of-time-ordered correlators},}\ }\href {\doibase
  10.1103/PhysRevE.98.012216} {\bibfield  {journal} {\bibinfo  {journal} {Phys.
  Rev. E}\ }\textbf {\bibinfo {volume} {98}},\ \bibinfo {pages} {012216}
  (\bibinfo {year} {2018})}\BibitemShut {NoStop}%
\bibitem [{\citenamefont {Swingle}\ and\ \citenamefont
  {Chowdhury}(2017)}]{Swingle2017-sj}%
  \BibitemOpen
  \bibfield  {author} {\bibinfo {author} {\bibfnamefont {B.}~\bibnamefont
  {Swingle}}\ and\ \bibinfo {author} {\bibfnamefont {D.}~\bibnamefont
  {Chowdhury}},\ }\bibfield  {title} {\enquote {\bibinfo {title} {Slow
  scrambling in disordered quantum systems},}\ }\href {\doibase
  10.1103/PhysRevB.95.060201} {\bibfield  {journal} {\bibinfo  {journal} {Phys.
  Rev. B}\ }\textbf {\bibinfo {volume} {95}},\ \bibinfo {pages} {060201}
  (\bibinfo {year} {2017})}\BibitemShut {NoStop}%
\bibitem [{\citenamefont {Campisi}\ and\ \citenamefont
  {Goold}(2017)}]{Campisi2017-ev}%
  \BibitemOpen
  \bibfield  {author} {\bibinfo {author} {\bibfnamefont {M.}~\bibnamefont
  {Campisi}}\ and\ \bibinfo {author} {\bibfnamefont {J.}~\bibnamefont
  {Goold}},\ }\bibfield  {title} {\enquote {\bibinfo {title} {Thermodynamics of
  quantum information scrambling},}\ }\href {\doibase
  10.1103/PhysRevE.95.062127} {\bibfield  {journal} {\bibinfo  {journal} {Phys.
  Rev. E}\ }\textbf {\bibinfo {volume} {95}},\ \bibinfo {pages} {062127}
  (\bibinfo {year} {2017})}\BibitemShut {NoStop}%
\bibitem [{\citenamefont {Chen}\ \emph {et~al.}(2017)\citenamefont {Chen},
  \citenamefont {Zhou}, \citenamefont {Huse},\ and\ \citenamefont
  {Fradkin}}]{Chen2017-yh}%
  \BibitemOpen
  \bibfield  {author} {\bibinfo {author} {\bibfnamefont {X.}~\bibnamefont
  {Chen}}, \bibinfo {author} {\bibfnamefont {T.}~\bibnamefont {Zhou}}, \bibinfo
  {author} {\bibfnamefont {D.~A.}\ \bibnamefont {Huse}}, \ and\ \bibinfo
  {author} {\bibfnamefont {E.}~\bibnamefont {Fradkin}},\ }\bibfield  {title}
  {\enquote {\bibinfo {title} {Out-of-time-order correlations in many-body
  localized and thermal phases},}\ }\href {\doibase 10.1002/andp.201600332}
  {\bibfield  {journal} {\bibinfo  {journal} {Annalen der Physik}\ }\textbf
  {\bibinfo {volume} {529}},\ \bibinfo {pages} {1600332} (\bibinfo {year}
  {2017})}\BibitemShut {NoStop}%
\bibitem [{\citenamefont {Rozenbaum}\ \emph {et~al.}(2017)\citenamefont
  {Rozenbaum}, \citenamefont {Ganeshan},\ and\ \citenamefont
  {Galitski}}]{Rozenbaum2017-fh}%
  \BibitemOpen
  \bibfield  {author} {\bibinfo {author} {\bibfnamefont {E.~B.}\ \bibnamefont
  {Rozenbaum}}, \bibinfo {author} {\bibfnamefont {S.}~\bibnamefont {Ganeshan}},
  \ and\ \bibinfo {author} {\bibfnamefont {V.}~\bibnamefont {Galitski}},\
  }\bibfield  {title} {\enquote {\bibinfo {title} {Lyapunov exponent and
  {Out-of-Time-Ordered} correlator's growth rate in a chaotic system},}\ }\href
  {\doibase 10.1103/PhysRevLett.118.086801} {\bibfield  {journal} {\bibinfo
  {journal} {Phys. Rev. Lett.}\ }\textbf {\bibinfo {volume} {118}},\ \bibinfo
  {pages} {086801} (\bibinfo {year} {2017})}\BibitemShut {NoStop}%
\bibitem [{\citenamefont {D{\'o}ra}\ and\ \citenamefont
  {Moessner}(2017)}]{Dora2017-mf}%
  \BibitemOpen
  \bibfield  {author} {\bibinfo {author} {\bibfnamefont {B.}~\bibnamefont
  {D{\'o}ra}}\ and\ \bibinfo {author} {\bibfnamefont {R.}~\bibnamefont
  {Moessner}},\ }\bibfield  {title} {\enquote {\bibinfo {title}
  {{Out-of-Time-Ordered Density Correlators in Luttinger Liquids}},}\ }\href
  {\doibase 10.1103/PhysRevLett.119.026802} {\bibfield  {journal} {\bibinfo
  {journal} {Phys. Rev. Lett.}\ }\textbf {\bibinfo {volume} {119}},\ \bibinfo
  {pages} {026802} (\bibinfo {year} {2017})}\BibitemShut {NoStop}%
\bibitem [{\citenamefont {Hashimoto}\ \emph {et~al.}(2017)\citenamefont
  {Hashimoto}, \citenamefont {Murata},\ and\ \citenamefont
  {Yoshii}}]{Hashimoto2017-ug}%
  \BibitemOpen
  \bibfield  {author} {\bibinfo {author} {\bibfnamefont {K.}~\bibnamefont
  {Hashimoto}}, \bibinfo {author} {\bibfnamefont {K.}~\bibnamefont {Murata}}, \
  and\ \bibinfo {author} {\bibfnamefont {R.}~\bibnamefont {Yoshii}},\
  }\bibfield  {title} {\enquote {\bibinfo {title} {Out-of-time-order
  correlators in quantum mechanics},}\ }\href {\doibase
  10.1007/JHEP10(2017)138} {\bibfield  {journal} {\bibinfo  {journal} {J. High
  Energy Phys}\ }\textbf {\bibinfo {volume} {10}},\ \bibinfo {pages} {138}
  (\bibinfo {year} {2017})}\BibitemShut {NoStop}%
\bibitem [{\citenamefont {Lin}\ and\ \citenamefont
  {Motrunich}(2018{\natexlab{a}})}]{Lin2018-mm}%
  \BibitemOpen
  \bibfield  {author} {\bibinfo {author} {\bibfnamefont {C.-J.}\ \bibnamefont
  {Lin}}\ and\ \bibinfo {author} {\bibfnamefont {O.~I.}\ \bibnamefont
  {Motrunich}},\ }\bibfield  {title} {\enquote {\bibinfo {title}
  {{Out-of-time-ordered correlators in a quantum Ising chain}},}\ }\href
  {\doibase 10.1103/PhysRevB.97.144304} {\bibfield  {journal} {\bibinfo
  {journal} {Phys. Rev. B}\ }\textbf {\bibinfo {volume} {97}},\ \bibinfo
  {pages} {144304} (\bibinfo {year} {2018}{\natexlab{a}})}\BibitemShut
  {NoStop}%
\bibitem [{\citenamefont {Lin}\ and\ \citenamefont
  {Motrunich}(2018{\natexlab{b}})}]{Lin2018-px}%
  \BibitemOpen
  \bibfield  {author} {\bibinfo {author} {\bibfnamefont {C.~J.}\ \bibnamefont
  {Lin}}\ and\ \bibinfo {author} {\bibfnamefont {O.~I.}\ \bibnamefont
  {Motrunich}},\ }\bibfield  {title} {\enquote {\bibinfo {title}
  {{Out-of-time-ordered correlators in short-range and long-range hard-core
  boson models and in the Luttinger-liquid model}},}\ }\href {\doibase
  10.1103/PhysRevB.98.134305} {\bibfield  {journal} {\bibinfo  {journal} {Phys.
  Rev. B}\ }\textbf {\bibinfo {volume} {98}},\ \bibinfo {pages} {134305}
  (\bibinfo {year} {2018}{\natexlab{b}})}\BibitemShut {NoStop}%
\bibitem [{\citenamefont {von Keyserlingk}\ \emph {et~al.}(2018)\citenamefont
  {von Keyserlingk}, \citenamefont {Rakovszky}, \citenamefont {Pollmann},\ and\
  \citenamefont {Sondhi}}]{Von_Keyserlingk2018-cf}%
  \BibitemOpen
  \bibfield  {author} {\bibinfo {author} {\bibfnamefont {C.~W.}\ \bibnamefont
  {von Keyserlingk}}, \bibinfo {author} {\bibfnamefont {T.}~\bibnamefont
  {Rakovszky}}, \bibinfo {author} {\bibfnamefont {F.}~\bibnamefont {Pollmann}},
  \ and\ \bibinfo {author} {\bibfnamefont {S.~L.}\ \bibnamefont {Sondhi}},\
  }\bibfield  {title} {\enquote {\bibinfo {title} {Operator hydrodynamics,
  {OTOCs}, and entanglement growth in systems without conservation laws},}\
  }\href {\doibase 10.1103/PhysRevX.8.021013} {\bibfield  {journal} {\bibinfo
  {journal} {Phys. Rev. X}\ }\textbf {\bibinfo {volume} {8}},\ \bibinfo {pages}
  {021013} (\bibinfo {year} {2018})}\BibitemShut {NoStop}%
\bibitem [{\citenamefont {Nahum}\ \emph {et~al.}(2018)\citenamefont {Nahum},
  \citenamefont {Vijay},\ and\ \citenamefont {Haah}}]{Nahum2018-ta}%
  \BibitemOpen
  \bibfield  {author} {\bibinfo {author} {\bibfnamefont {A.}~\bibnamefont
  {Nahum}}, \bibinfo {author} {\bibfnamefont {S.}~\bibnamefont {Vijay}}, \ and\
  \bibinfo {author} {\bibfnamefont {J.}~\bibnamefont {Haah}},\ }\bibfield
  {title} {\enquote {\bibinfo {title} {Operator spreading in random unitary
  circuits},}\ }\href {\doibase 10.1103/PhysRevX.8.021014} {\bibfield
  {journal} {\bibinfo  {journal} {Phys. Rev. X}\ }\textbf {\bibinfo {volume}
  {8}},\ \bibinfo {pages} {021014} (\bibinfo {year} {2018})}\BibitemShut
  {NoStop}%
\bibitem [{\citenamefont {Khemani}\ \emph {et~al.}(2018)\citenamefont
  {Khemani}, \citenamefont {Vishwanath},\ and\ \citenamefont
  {Huse}}]{Khemani2018-cu}%
  \BibitemOpen
  \bibfield  {author} {\bibinfo {author} {\bibfnamefont {V.}~\bibnamefont
  {Khemani}}, \bibinfo {author} {\bibfnamefont {A.}~\bibnamefont {Vishwanath}},
  \ and\ \bibinfo {author} {\bibfnamefont {D.~A.}\ \bibnamefont {Huse}},\
  }\bibfield  {title} {\enquote {\bibinfo {title} {Operator spreading and the
  emergence of dissipative hydrodynamics under unitary evolution with
  conservation laws},}\ }\href {\doibase 10.1103/PhysRevX.8.031057} {\bibfield
  {journal} {\bibinfo  {journal} {Phys. Rev. X}\ }\textbf {\bibinfo {volume}
  {8}},\ \bibinfo {pages} {031057} (\bibinfo {year} {2018})}\BibitemShut
  {NoStop}%
\bibitem [{\citenamefont {Gharibyan}\ \emph {et~al.}(2018)\citenamefont
  {Gharibyan}, \citenamefont {Hanada}, \citenamefont {Shenker},\ and\
  \citenamefont {Tezuka}}]{Gharibyan2018-fe}%
  \BibitemOpen
  \bibfield  {author} {\bibinfo {author} {\bibfnamefont {H.}~\bibnamefont
  {Gharibyan}}, \bibinfo {author} {\bibfnamefont {M.}~\bibnamefont {Hanada}},
  \bibinfo {author} {\bibfnamefont {S.~H.}\ \bibnamefont {Shenker}}, \ and\
  \bibinfo {author} {\bibfnamefont {M.}~\bibnamefont {Tezuka}},\ }\bibfield
  {title} {\enquote {\bibinfo {title} {Onset of random matrix behavior in
  scrambling systems},}\ }\href {\doibase 10.1007/JHEP07(2018)124} {\bibfield
  {journal} {\bibinfo  {journal} {J. High Energy Phys.}\ }\textbf {\bibinfo
  {volume} {07}},\ \bibinfo {pages} {124} (\bibinfo {year} {2018})}\BibitemShut
  {NoStop}%
\bibitem [{\citenamefont {Xu}\ and\ \citenamefont {Swingle}(2019)}]{Xu2018-hw}%
  \BibitemOpen
  \bibfield  {author} {\bibinfo {author} {\bibfnamefont {Shenglong}\
  \bibnamefont {Xu}}\ and\ \bibinfo {author} {\bibfnamefont {Brian}\
  \bibnamefont {Swingle}},\ }\bibfield  {title} {\enquote {\bibinfo {title}
  {Locality, quantum fluctuations, and scrambling},}\ }\href {\doibase
  10.1103/PhysRevX.9.031048} {\bibfield  {journal} {\bibinfo  {journal} {Phys.
  Rev. X}\ }\textbf {\bibinfo {volume} {9}},\ \bibinfo {pages} {031048}
  (\bibinfo {year} {2019})}\BibitemShut {NoStop}%
\bibitem [{\citenamefont {Tuziemski}(2019)}]{Tuziemski2019-kq}%
  \BibitemOpen
  \bibfield  {author} {\bibinfo {author} {\bibfnamefont {Jan}\ \bibnamefont
  {Tuziemski}},\ }\bibfield  {title} {\enquote {\bibinfo {title}
  {Out-of-time-ordered correlation functions in open systems: A
  {Feynman-Vernon} influence functional approach},}\ }\href {\doibase
  10.1103/PhysRevA.100.062106} {\bibfield  {journal} {\bibinfo  {journal}
  {Phys. Rev. A}\ }\textbf {\bibinfo {volume} {100}},\ \bibinfo {pages}
  {062106} (\bibinfo {year} {2019})}\BibitemShut {NoStop}%
\bibitem [{\citenamefont {Swingle}\ \emph {et~al.}(2016)\citenamefont
  {Swingle}, \citenamefont {Bentsen}, \citenamefont {Schleier-Smith},\ and\
  \citenamefont {Hayden}}]{Swingle2016-hh}%
  \BibitemOpen
  \bibfield  {author} {\bibinfo {author} {\bibfnamefont {B.}~\bibnamefont
  {Swingle}}, \bibinfo {author} {\bibfnamefont {G.}~\bibnamefont {Bentsen}},
  \bibinfo {author} {\bibfnamefont {M.}~\bibnamefont {Schleier-Smith}}, \ and\
  \bibinfo {author} {\bibfnamefont {P.}~\bibnamefont {Hayden}},\ }\bibfield
  {title} {\enquote {\bibinfo {title} {Measuring the scrambling of quantum
  information},}\ }\href {\doibase 10.1103/PhysRevA.94.040302} {\bibfield
  {journal} {\bibinfo  {journal} {Phys. Rev. A}\ }\textbf {\bibinfo {volume}
  {94}},\ \bibinfo {pages} {040302(R)} (\bibinfo {year} {2016})}\BibitemShut
  {NoStop}%
\bibitem [{\citenamefont {G{\"a}rttner}\ \emph {et~al.}(2017)\citenamefont
  {G{\"a}rttner}, \citenamefont {Bohnet}, \citenamefont {Safavi-Naini},
  \citenamefont {Wall}, \citenamefont {Bollinger},\ and\ \citenamefont
  {Rey}}]{Garttner2017-rl}%
  \BibitemOpen
  \bibfield  {author} {\bibinfo {author} {\bibfnamefont {M.}~\bibnamefont
  {G{\"a}rttner}}, \bibinfo {author} {\bibfnamefont {J.~G.}\ \bibnamefont
  {Bohnet}}, \bibinfo {author} {\bibfnamefont {A.}~\bibnamefont
  {Safavi-Naini}}, \bibinfo {author} {\bibfnamefont {M.~L.}\ \bibnamefont
  {Wall}}, \bibinfo {author} {\bibfnamefont {J.~J.}\ \bibnamefont {Bollinger}},
  \ and\ \bibinfo {author} {\bibfnamefont {A.~M.}\ \bibnamefont {Rey}},\
  }\bibfield  {title} {\enquote {\bibinfo {title} {Measuring out-of-time-order
  correlations and multiple quantum spectra in a trapped-ion quantum magnet},}\
  }\href {\doibase 10.1038/nphys4119} {\bibfield  {journal} {\bibinfo
  {journal} {Nat. Phys.}\ }\textbf {\bibinfo {volume} {13}},\ \bibinfo {pages}
  {781} (\bibinfo {year} {2017})}\BibitemShut {NoStop}%
\bibitem [{\citenamefont {Li}\ \emph {et~al.}(2017)\citenamefont {Li},
  \citenamefont {Fan}, \citenamefont {Wang}, \citenamefont {Ye}, \citenamefont
  {Zeng}, \citenamefont {Zhai}, \citenamefont {Peng},\ and\ \citenamefont
  {Du}}]{Li2017-il}%
  \BibitemOpen
  \bibfield  {author} {\bibinfo {author} {\bibfnamefont {J.}~\bibnamefont
  {Li}}, \bibinfo {author} {\bibfnamefont {R.}~\bibnamefont {Fan}}, \bibinfo
  {author} {\bibfnamefont {H.}~\bibnamefont {Wang}}, \bibinfo {author}
  {\bibfnamefont {B.}~\bibnamefont {Ye}}, \bibinfo {author} {\bibfnamefont
  {B.}~\bibnamefont {Zeng}}, \bibinfo {author} {\bibfnamefont {H.}~\bibnamefont
  {Zhai}}, \bibinfo {author} {\bibfnamefont {X.}~\bibnamefont {Peng}}, \ and\
  \bibinfo {author} {\bibfnamefont {J.}~\bibnamefont {Du}},\ }\bibfield
  {title} {\enquote {\bibinfo {title} {Measuring {Out-of-Time-Order}
  correlators on a nuclear magnetic resonance quantum simulator},}\ }\href
  {\doibase 10.1103/PhysRevX.7.031011} {\bibfield  {journal} {\bibinfo
  {journal} {Phys. Rev. X}\ }\textbf {\bibinfo {volume} {7}},\ \bibinfo {pages}
  {031011} (\bibinfo {year} {2017})}\BibitemShut {NoStop}%
\bibitem [{\citenamefont {Landsman}\ \emph {et~al.}(2019)\citenamefont
  {Landsman}, \citenamefont {Figgatt}, \citenamefont {Schuster}, \citenamefont
  {Linke}, \citenamefont {Yoshida}, \citenamefont {Yao},\ and\ \citenamefont
  {Monroe}}]{Landsman2019-lf}%
  \BibitemOpen
  \bibfield  {author} {\bibinfo {author} {\bibfnamefont {K~A}\ \bibnamefont
  {Landsman}}, \bibinfo {author} {\bibfnamefont {C}~\bibnamefont {Figgatt}},
  \bibinfo {author} {\bibfnamefont {T}~\bibnamefont {Schuster}}, \bibinfo
  {author} {\bibfnamefont {N~M}\ \bibnamefont {Linke}}, \bibinfo {author}
  {\bibfnamefont {B}~\bibnamefont {Yoshida}}, \bibinfo {author} {\bibfnamefont
  {N~Y}\ \bibnamefont {Yao}}, \ and\ \bibinfo {author} {\bibfnamefont
  {C}~\bibnamefont {Monroe}},\ }\bibfield  {title} {\enquote {\bibinfo {title}
  {Verified quantum information scrambling},}\ }\href {\doibase
  10.1038/s41586-019-0952-6} {\bibfield  {journal} {\bibinfo  {journal}
  {Nature}\ }\textbf {\bibinfo {volume} {567}},\ \bibinfo {pages} {61--65}
  (\bibinfo {year} {2019})},\ \Eprint {http://arxiv.org/abs/1806.02807}
  {arXiv:1806.02807 [quant-ph]} \BibitemShut {NoStop}%
\bibitem [{\citenamefont {Kurchan}(2018)}]{Kurchan2018-ff}%
  \BibitemOpen
  \bibfield  {author} {\bibinfo {author} {\bibfnamefont {J.}~\bibnamefont
  {Kurchan}},\ }\bibfield  {title} {\enquote {\bibinfo {title} {Quantum bound
  to chaos and the semiclassical limit},}\ }\href {\doibase
  10.1007/s10955-018-2052-7} {\bibfield  {journal} {\bibinfo  {journal} {J.
  Stat. Phys.}\ }\textbf {\bibinfo {volume} {171}},\ \bibinfo {pages}
  {965--979} (\bibinfo {year} {2018})}\BibitemShut {NoStop}%
\bibitem [{\citenamefont {Schmitt}\ \emph {et~al.}(2019)\citenamefont
  {Schmitt}, \citenamefont {Sels}, \citenamefont {Kehrein},\ and\ \citenamefont
  {Polkovnikov}}]{Schmitt2019-gg}%
  \BibitemOpen
  \bibfield  {author} {\bibinfo {author} {\bibfnamefont {M.}~\bibnamefont
  {Schmitt}}, \bibinfo {author} {\bibfnamefont {D.}~\bibnamefont {Sels}},
  \bibinfo {author} {\bibfnamefont {S.}~\bibnamefont {Kehrein}}, \ and\
  \bibinfo {author} {\bibfnamefont {A.}~\bibnamefont {Polkovnikov}},\
  }\bibfield  {title} {\enquote {\bibinfo {title} {Semiclassical echo dynamics
  in the {Sachdev-Ye-Kitaev} model},}\ }\href {\doibase
  10.1103/PhysRevB.99.134301} {\bibfield  {journal} {\bibinfo  {journal} {Phys.
  Rev. B}\ }\textbf {\bibinfo {volume} {99}},\ \bibinfo {pages} {134301}
  (\bibinfo {year} {2019})}\BibitemShut {NoStop}%
\bibitem [{\citenamefont {Hamazaki}\ \emph {et~al.}(2018)\citenamefont
  {Hamazaki}, \citenamefont {Fujimoto},\ and\ \citenamefont
  {Ueda}}]{Hamazaki2018-ep}%
  \BibitemOpen
  \bibfield  {author} {\bibinfo {author} {\bibfnamefont {R.}~\bibnamefont
  {Hamazaki}}, \bibinfo {author} {\bibfnamefont {K.}~\bibnamefont {Fujimoto}},
  \ and\ \bibinfo {author} {\bibfnamefont {M.}~\bibnamefont {Ueda}},\
  }\bibfield  {title} {\enquote {\bibinfo {title} {Operator noncommutativity
  and irreversibility in quantum chaos},}\ }\href
  {http://arxiv.org/abs/1807.02360} {\  (\bibinfo {year} {2018})},\ \Eprint
  {http://arxiv.org/abs/1807.02360} {arXiv:1807.02360 [cond-mat.stat-mech]}
  \BibitemShut {NoStop}%
\bibitem [{\citenamefont {Fan}\ \emph {et~al.}(2017)\citenamefont {Fan},
  \citenamefont {Zhang}, \citenamefont {Shen},\ and\ \citenamefont
  {Zhai}}]{Fan2017-vy}%
  \BibitemOpen
  \bibfield  {author} {\bibinfo {author} {\bibfnamefont {Ruihua}\ \bibnamefont
  {Fan}}, \bibinfo {author} {\bibfnamefont {Pengfei}\ \bibnamefont {Zhang}},
  \bibinfo {author} {\bibfnamefont {Huitao}\ \bibnamefont {Shen}}, \ and\
  \bibinfo {author} {\bibfnamefont {Hui}\ \bibnamefont {Zhai}},\ }\bibfield
  {title} {\enquote {\bibinfo {title} {Out-of-time-order correlation for
  many-body localization},}\ }\href {\doibase 10.1016/j.scib.2017.04.011}
  {\bibfield  {journal} {\bibinfo  {journal} {Science Bulletin of the Faculty
  of Agriculture, Kyushu University}\ }\textbf {\bibinfo {volume} {62}},\
  \bibinfo {pages} {707--711} (\bibinfo {year} {2017})}\BibitemShut {NoStop}%
\bibitem [{\citenamefont {Cotler}\ \emph
  {et~al.}(2017{\natexlab{a}})\citenamefont {Cotler}, \citenamefont
  {Hunter-Jones}, \citenamefont {Liu},\ and\ \citenamefont
  {Yoshida}}]{Cotler2017-oi}%
  \BibitemOpen
  \bibfield  {author} {\bibinfo {author} {\bibfnamefont {J.}~\bibnamefont
  {Cotler}}, \bibinfo {author} {\bibfnamefont {N.}~\bibnamefont
  {Hunter-Jones}}, \bibinfo {author} {\bibfnamefont {J.}~\bibnamefont {Liu}}, \
  and\ \bibinfo {author} {\bibfnamefont {B.}~\bibnamefont {Yoshida}},\
  }\bibfield  {title} {\enquote {\bibinfo {title} {Chaos, complexity, and
  random matrices},}\ }\href {\doibase 10.1007/JHEP11(2017)048} {\bibfield
  {journal} {\bibinfo  {journal} {J. High Energy Phys.}\ }\textbf {\bibinfo
  {volume} {11}},\ \bibinfo {pages} {48} (\bibinfo {year}
  {2017}{\natexlab{a}})}\BibitemShut {NoStop}%
\bibitem [{\citenamefont {de~Mello~Koch}\ \emph {et~al.}(2019)\citenamefont
  {de~Mello~Koch}, \citenamefont {Huang}, \citenamefont {Ma},\ and\
  \citenamefont {Van~Zyl}}]{De_Mello_Koch2019-ez}%
  \BibitemOpen
  \bibfield  {author} {\bibinfo {author} {\bibfnamefont {R.}~\bibnamefont
  {de~Mello~Koch}}, \bibinfo {author} {\bibfnamefont {J-H}\ \bibnamefont
  {Huang}}, \bibinfo {author} {\bibfnamefont {C-T}\ \bibnamefont {Ma}}, \ and\
  \bibinfo {author} {\bibfnamefont {H.~J.~R.}\ \bibnamefont {Van~Zyl}},\
  }\bibfield  {title} {\enquote {\bibinfo {title} {Spectral form factor as an
  {OTOC} averaged over the heisenberg group},}\ }\href {\doibase
  10.1016/j.physletb.2019.06.025} {\bibfield  {journal} {\bibinfo  {journal}
  {Phys. Lett. B}\ }\textbf {\bibinfo {volume} {795}},\ \bibinfo {pages}
  {183--187} (\bibinfo {year} {2019})}\BibitemShut {NoStop}%
\bibitem [{\citenamefont {Ma}(2019)}]{Ma2019-sw}%
  \BibitemOpen
  \bibfield  {author} {\bibinfo {author} {\bibfnamefont {C-T}\ \bibnamefont
  {Ma}},\ }\bibfield  {title} {\enquote {\bibinfo {title} {{Early-Time} and
  {Late-Time} quantum chaos},}\ }\href {http://arxiv.org/abs/1907.04289} {\
  (\bibinfo {year} {2019})},\ \Eprint {http://arxiv.org/abs/1907.04289}
  {arXiv:1907.04289 [hep-th]} \BibitemShut {NoStop}%
\bibitem [{\citenamefont {Cotler}\ \emph
  {et~al.}(2017{\natexlab{b}})\citenamefont {Cotler}, \citenamefont {Gur-Ari},
  \citenamefont {Hanada}, \citenamefont {Polchinski}, \citenamefont {{Saa}},
  \citenamefont {Shenker}, \citenamefont {Stanford}, \citenamefont
  {Streicher},\ and\ \citenamefont {Tezuka}}]{Cotler2017-dq}%
  \BibitemOpen
  \bibfield  {author} {\bibinfo {author} {\bibfnamefont {J.}~\bibnamefont
  {Cotler}}, \bibinfo {author} {\bibfnamefont {G.}~\bibnamefont {Gur-Ari}},
  \bibinfo {author} {\bibfnamefont {M.}~\bibnamefont {Hanada}}, \bibinfo
  {author} {\bibfnamefont {J.}~\bibnamefont {Polchinski}}, \bibinfo {author}
  {\bibnamefont {{Saa}}}, \bibinfo {author} {\bibfnamefont {S.~H.}\
  \bibnamefont {Shenker}}, \bibinfo {author} {\bibfnamefont {D.}~\bibnamefont
  {Stanford}}, \bibinfo {author} {\bibfnamefont {A.}~\bibnamefont {Streicher}},
  \ and\ \bibinfo {author} {\bibfnamefont {M.}~\bibnamefont {Tezuka}},\
  }\bibfield  {title} {\enquote {\bibinfo {title} {Black holes and random
  matrices},}\ }\href {\doibase 10.1007/JHEP05(2017)118} {\bibfield  {journal}
  {\bibinfo  {journal} {J. High Energy Phys.}\ }\textbf {\bibinfo {volume}
  {05}},\ \bibinfo {pages} {118} (\bibinfo {year}
  {2017}{\natexlab{b}})}\BibitemShut {NoStop}%
\bibitem [{\citenamefont {Bhattacharyya}\ \emph {et~al.}(2019)\citenamefont
  {Bhattacharyya}, \citenamefont {Chemissany}, \citenamefont {Shajidul~Haque},\
  and\ \citenamefont {Yan}}]{Bhattacharyya2019-nx}%
  \BibitemOpen
  \bibfield  {author} {\bibinfo {author} {\bibfnamefont {A.}~\bibnamefont
  {Bhattacharyya}}, \bibinfo {author} {\bibfnamefont {W.}~\bibnamefont
  {Chemissany}}, \bibinfo {author} {\bibfnamefont {S.}~\bibnamefont
  {Shajidul~Haque}}, \ and\ \bibinfo {author} {\bibfnamefont {B.}~\bibnamefont
  {Yan}},\ }\bibfield  {title} {\enquote {\bibinfo {title} {Towards the web of
  quantum chaos diagnostics},}\ }\href {http://arxiv.org/abs/1909.01894} {\
  (\bibinfo {year} {2019})},\ \Eprint {http://arxiv.org/abs/1909.01894}
  {arXiv:1909.01894 [hep-th]} \BibitemShut {NoStop}%
\bibitem [{\citenamefont {Zurek}(2003)}]{Zurek2003-qi}%
  \BibitemOpen
  \bibfield  {author} {\bibinfo {author} {\bibfnamefont {W.~H.}\ \bibnamefont
  {Zurek}},\ }\bibfield  {title} {\enquote {\bibinfo {title} {Decoherence,
  einselection, and the quantum origins of the classical},}\ }\href {\doibase
  10.1103/RevModPhys.75.715} {\bibfield  {journal} {\bibinfo  {journal} {Rev.
  Mod. Phys.}\ }\textbf {\bibinfo {volume} {75}},\ \bibinfo {pages} {715--775}
  (\bibinfo {year} {2003})}\BibitemShut {NoStop}%
\bibitem [{\citenamefont {Schlosshauer}(2007)}]{Schlosshauer2007-kp}%
  \BibitemOpen
  \bibfield  {author} {\bibinfo {author} {\bibfnamefont {M.}~\bibnamefont
  {Schlosshauer}},\ }\href {\doibase 10.1007/978-3-540-35775-9} {\emph
  {\bibinfo {title} {Decoherence and the {Quantum-To-Classical} Transition}}}\
  (\bibinfo  {publisher} {Springer, Berlin, Heidelberg},\ \bibinfo {year}
  {2007})\BibitemShut {NoStop}%
\bibitem [{\citenamefont {Budini}\ \emph {et~al.}(1999)\citenamefont {Budini},
  \citenamefont {Karina~Chattah},\ and\ \citenamefont
  {C{\'a}ceres}}]{Budini1999-uo}%
  \BibitemOpen
  \bibfield  {author} {\bibinfo {author} {\bibfnamefont {A.~A.}\ \bibnamefont
  {Budini}}, \bibinfo {author} {\bibfnamefont {A.}~\bibnamefont
  {Karina~Chattah}}, \ and\ \bibinfo {author} {\bibfnamefont {M.~O.}\
  \bibnamefont {C{\'a}ceres}},\ }\bibfield  {title} {\enquote {\bibinfo {title}
  {On the quantum dissipative generator: weak-coupling approximation and
  stochastic approach},}\ }\href {\doibase 10.1088/0305-4470/32/4/007}
  {\bibfield  {journal} {\bibinfo  {journal} {J. Phys. A}\ }\textbf {\bibinfo
  {volume} {32}},\ \bibinfo {pages} {631} (\bibinfo {year} {1999})}\BibitemShut
  {NoStop}%
\bibitem [{\citenamefont {Romero-Berm{\'u}dez}\ \emph
  {et~al.}(2019)\citenamefont {Romero-Berm{\'u}dez}, \citenamefont {Schalm},\
  and\ \citenamefont {Scopelliti}}]{Romero-Bermudez2019-lj}%
  \BibitemOpen
  \bibfield  {author} {\bibinfo {author} {\bibfnamefont {A.}~\bibnamefont
  {Romero-Berm{\'u}dez}}, \bibinfo {author} {\bibfnamefont {K.}~\bibnamefont
  {Schalm}}, \ and\ \bibinfo {author} {\bibfnamefont {V.}~\bibnamefont
  {Scopelliti}},\ }\bibfield  {title} {\enquote {\bibinfo {title}
  {Regularization dependence of the {OTOC}. which lyapunov spectrum is the
  physical one?}}\ }\href {http://arxiv.org/abs/1903.09595} {\  (\bibinfo
  {year} {2019})},\ \Eprint {http://arxiv.org/abs/1903.09595} {arXiv:1903.09595
  [hep-th]} \BibitemShut {NoStop}%
\bibitem [{\citenamefont {Blume-Kohout}\ and\ \citenamefont
  {Zurek}(2003)}]{Blume-Kohout2003-ww}%
  \BibitemOpen
  \bibfield  {author} {\bibinfo {author} {\bibfnamefont {R.}~\bibnamefont
  {Blume-Kohout}}\ and\ \bibinfo {author} {\bibfnamefont {W.~H.}\ \bibnamefont
  {Zurek}},\ }\bibfield  {title} {\enquote {\bibinfo {title} {Decoherence from
  a chaotic environment: An upside-down ``oscillator'' as a model},}\ }\href
  {\doibase 10.1103/PhysRevA.68.032104} {\bibfield  {journal} {\bibinfo
  {journal} {Phys. Rev. A}\ }\textbf {\bibinfo {volume} {68}},\ \bibinfo
  {pages} {032104} (\bibinfo {year} {2003})}\BibitemShut {NoStop}%
\bibitem [{\citenamefont {Xu}\ and\ \citenamefont {Swingle}(2020)}]{Xu2020-xo}%
  \BibitemOpen
  \bibfield  {author} {\bibinfo {author} {\bibfnamefont {Shenglong}\
  \bibnamefont {Xu}}\ and\ \bibinfo {author} {\bibfnamefont {Brian}\
  \bibnamefont {Swingle}},\ }\bibfield  {title} {\enquote {\bibinfo {title}
  {Accessing scrambling using matrix product operators},}\ }\href {\doibase
  10.1038/s41567-019-0712-4} {\bibfield  {journal} {\bibinfo  {journal} {Nat.
  Phys.}\ }\textbf {\bibinfo {volume} {16}},\ \bibinfo {pages} {199--204}
  (\bibinfo {year} {2020})}\BibitemShut {NoStop}%
\bibitem [{\citenamefont {Sachdev}\ and\ \citenamefont
  {Ye}(1993)}]{Sachdev1993-ka}%
  \BibitemOpen
  \bibfield  {author} {\bibinfo {author} {\bibfnamefont {S.}~\bibnamefont
  {Sachdev}}\ and\ \bibinfo {author} {\bibfnamefont {J.}~\bibnamefont {Ye}},\
  }\bibfield  {title} {\enquote {\bibinfo {title} {{Gapless spin-fluid ground
  state in a random quantum Heisenberg magnet}},}\ }\href {\doibase
  10.1103/PhysRevLett.70.3339} {\bibfield  {journal} {\bibinfo  {journal}
  {Phys. Rev. Lett.}\ }\textbf {\bibinfo {volume} {70}},\ \bibinfo {pages}
  {3339--3342} (\bibinfo {year} {1993})}\BibitemShut {NoStop}%
\bibitem [{\citenamefont {Maldacena}\ and\ \citenamefont
  {Stanford}(2016)}]{Maldacena2016-yh}%
  \BibitemOpen
  \bibfield  {author} {\bibinfo {author} {\bibfnamefont {J.}~\bibnamefont
  {Maldacena}}\ and\ \bibinfo {author} {\bibfnamefont {D.}~\bibnamefont
  {Stanford}},\ }\bibfield  {title} {\enquote {\bibinfo {title} {Remarks on the
  {Sachdev-Ye-Kitaev} model},}\ }\href {\doibase 10.1103/PhysRevD.94.106002}
  {\bibfield  {journal} {\bibinfo  {journal} {Phys. Rev. D}\ }\textbf {\bibinfo
  {volume} {94}},\ \bibinfo {pages} {106002} (\bibinfo {year}
  {2016})}\BibitemShut {NoStop}%
\end{thebibliography}%

\clearpage

\appendix

\setcounter{page}{1}
\renewcommand\thefigure{\thesection.\arabic{figure}}
\setcounter{figure}{0}

\onecolumngrid


\section{Haar integral of subsystem operators}\label{app:Haar}
In this appendix we prove the formula in Eq.~(\ref{eq:Haar}) in the main text. 

For a single system, the Haar integral over a unitary operator $A$ gives 
\begin{equation}\label{eq:Haarint}
    \int dA\ A^{\dagger}OA=\frac{1}{d}\Tr(O)\mathbb{I}. 
\end{equation}
Here, $dA\equiv d\mu(A)$ with $\mu$ the Haar measure on a unitary group.

This is a consequence of the defining property of the Haar measure, namely, the Haar measure is invariant under the transformation $A\rightarrow AU$, i.e., $\mu(A)=\mu(AU)$, where $U$ is an arbitrary unitary operator. This implies
\begin{equation}
    \int d\mu(A) A^{\dagger}OA=U^\dagger\left( \int d\mu(A) A^{\dagger}OA \right)U.
\end{equation}
Thus, the Haar average of a operator $A$ is proportional to the identity operator. In finite dimension, its trace can be computed as 
\begin{equation}
    \Tr \int d\mu(A) A^{\dagger}OA = \int d\mu(A) \Tr(O).
\end{equation}
The above integral is unique up to a constant multiplication factor; and the unitary groups on finite dimensional Hilbert spaces have finite measures. We can then choose $\int d\mu=1$ as a convention. The desired property in Eq.~(\ref{eq:Haarint}) is then deduced. We note that the identity $\int dA\ A^{\dagger}OA=\Tr(O)\mathbb{I}$ holds in infinite dimensions as well. The formal mathematical treatment of the infinite dimensional case will be published elsewhere \cite{Bhattacharyya2019-nx}.

For an operator $O_{AB}$ on a composed system, the integral over operators on a subsystem can be performed in a similar way, by performing the partial trace:
\begin{equation}
\begin{aligned}
       &\int dA\  A^{\dagger}\otimes\mathbb{I}_B O_{AB}A\otimes\mathbb{I}_B\\
       &=\int dA\ A^{\dagger}\otimes\mathbb{I}_B \sum_{i} O^A_i\otimes O^B_iA\otimes\mathbb{I}_B\\
       &=\sum_{i} \int dA\ A^{\dagger} O^A_i A \otimes O^B_i\\
       &=\frac{1}{d_A}\sum_i \Tr(O^A_i)\mathbb{I}_A \otimes O^B_i
       =\frac{1}{d_A}\mathbb{I}_A \otimes \Tr_{A}O_{AB} \ .
\end{aligned}
\end{equation}

\section{Reduced dynamics for operator $B$}\label{app:B}
This appendix presents a more rigorous derivation of the reduced dynamical equation for the operator $B$ in Eq.~(\ref{eq:bdynamics}) of the main text.

The total system Hamiltonian admits a general decomposition

\begin{equation}
\begin{aligned}
       & H=\mathbb{I}_A\otimes H_B + H_A\otimes \mathbb{I}_B + H',\\
    &  H'\equiv\delta\sum_{k=1}^{d^2_A}V_A^k\otimes V_B^k.
\end{aligned}
\end{equation}
where $A$ denotes a small local subsystem $S_A$, $B$ denotes the compliment of $S_A$ to the total system. It is assumed that the total system is much larger than the local system $S_A$. $d_A$ is the dimension of $S_A$. The operators $\{V_A^i\}_i$ are chosen to be Hermitian and orthonormal, with respect to the (weighted) Hilbert-Schmidt inner product, i.e.,
    \begin{equation}
        \Tr(V_A^iV_A^j)=d_A\delta_{i,j}.
    \end{equation}
Here, $d_A$ is introduced as a convention to normalize the noise correlation functions (see below). If the operators $\{V^i_A\}_i$ are not normalized in this way, proper normalization factors will appear in the stochastic field as pre-factors. The operators $V_B^i$ on $S_B$ are also Hermitian, but their (Hilbert-Schmidt) norms are fixed and are equal to the norms of $H_B$. Thus, the parameter $\delta$ quantifies the relative strength of the couplings compared with $H_B$.

We are interested in strongly coupled systems, where the energy scales admits a hierarchy $\bar{H}_A\ll\bar{H}_I\ll\bar{H}_B$. For instance, in an $N$-particle system with all-to-all two-body interactions, when the subsystem $S_A$ refers to a single particle, the energy scales of $S_A$, $S_B$, and the coupling between them, are of the order of $1$, $N^2$ and $N$, respectively. Alternatively speaking, this condition means that the parameter $\delta\ll1$; and energy scales of the subsystem $S_A$ is much smaller than that of the global dynamics.
    
Consider the reduced dynamics of an operator $B$ on the subsystem $S_B$ after the trace-out procedure. We have,
    \begin{equation}
    B(-t) = \Tr_A \left(e^{-i H t} \mathbb{I}_A\otimes B e^{i H t}\right). 
    \end{equation}
    This can be thought of as a decoherence process, i.e., the total system is prepared in an initial (unnormalized) product state $d_A\left(\frac{I_A}{d_A}\right)\otimes B$, where the subsystem $S_B$ is described by a ``density matrix'' $B$, and the subsystem $S_A$ is in a thermal state with infinite temperature up to normalization. The ``quantum state'' $B$ will become ``mixed'' with time evolution due to the presence of the couplings to subsystem $S_A$. When $\delta\ll 1$, the above evolution of $B(-t)$ can be expanded to the second order of $\delta$. This corresponds to the Born-Markov approximation, which leads the effective master equation for $B(-t)$ to a Lindblad form. It is known that in this case the effective master equation can be simulated with the evolution of $B$ under $H_B$ without coupling to other systems \cite{Budini1999-uo}, but subjects to a stochastic field
    \begin{equation}
        \delta \mathcal{F}(t)=\delta \sum_k l_k(t) V_B^k,
    \end{equation}
    with the correlations given by
    \begin{equation}
    \begin{aligned}
        &\overline{l_i(t)l_j(t-\tau)} \\
        = &\Tr(\frac{\mathbb{I}_A}{d_A}V_A^i e^{i H_A \tau}V^j_A e^{-i H_A \tau})\\
        \approx &\delta_{i,j}.
    \end{aligned}
    \end{equation}
    The approximation in the last step is due to the large energy hierarchy: the time scale of the dynamics of the subsystem $S_A$ is much larger than that of $B(t)$ under consideration. Alternatively, this can be thought of as taking the zeroth order the $H_A$. 
    As a consequence, the noise field $l_i(t)$ can be taken as a random constant $\pm 1$ with equal probability. The reduced dynamics of the $B$ operator is then given by
    \begin{equation}
        B(t)=d_A\times \overline{e^{-i (H_B + \mathcal{F})t } B e^{i (H_B + \mathcal{F})t }},
    \end{equation}
    averaged over the stochastic field $\mathcal{F}$. Note that the pre-factor $d_A$ is present in the above equation due to the normalization of $\mathbb{I}_A$. 
    
    As the noise fields are random $\pm 1$, each realization of the stochastic field, denoted as $V_\alpha$, always appears as linear combination of $V^i_B$'s, whose coefficients take values $\pm 1$ randomly. The noisy evolution of $B(-t)$ is then given by
\begin{equation}\label{eq:appendix_Boperator}
        B(-t) \approx d_A\times \frac{1}{N}\sum_{\alpha=1}^{N} e^{-i (H_B + \delta V_\alpha) t } B e^{i (H_B + \delta V_\alpha))t },
\end{equation}
where $N$ is the number of different realizations of the noise field.

\section{OTOC - LE connection} \label{app:C}

This appendix presents the derivation of Eq.~(\ref{eq:LE}) in the main text. It has been shown in the main text that after performing averaging over operator $A$, the OTOC reads
\begin{equation}
\begin{aligned}
       &\int dA\ F_{\beta=0}(t) \equiv \frac{1}{d} \int dA\ \Tr(A^\dagger(t) B^\dagger A(t) B)\\
       = &\frac{1}{d}\frac{1}{d_A} \Tr_B [ \Tr_A(e^{-iHt}B^\dagger e^{iHt})\Tr_A(e^{-iHt}Be^{iHt}) ].
\end{aligned}
\end{equation}
Replacing the reduced dynamics of operator $B$ with its alternative expression Eq.~(\ref{eq:appendix_Boperator}) allows us to further evaluate the average over all unitary operators on subsystem $S_B$:

\begin{equation}
\begin{aligned}
    &\overline{F_{\beta=0}(t)}\\
    =&\frac{1}{d}\frac{1}{d_A} \int dB\ \Tr_B [ \Tr_A(e^{-iHt}B^\dagger e^{iHt}) \Tr_A(e^{-iHt}Be^{iHt}) ]\\
    \approx & \frac{d_A}{d} \frac{1}{N^2}\sum_{\alpha,\alpha'=1} \Tr \int dB\  e^{-i(H_B+V_\alpha)t}B^\dagger e^{i(H_B+V_\alpha)t} \times \\
    &\ \ \ \ \ \ \ \ \ \ \ \ \ \  \times e^{-i(H_B+V_{\alpha'})t}Be^{i(H_B+V_{\alpha'})t} \\
    =&\frac{1}{N^2} \sum_{\alpha,\alpha'=1}\ |\frac{1}{d_B}\Tr(e^{i(H_B+V_\alpha)t}e^{-i(H_B+V_{\alpha'})t})|^2\\
    =&\overline{|\langle e^{i(H_B+V_\alpha) t}e^{-i(H_B+V_{\alpha'}) t}\rangle_{\beta=0}|^2}.
\end{aligned}
\end{equation}
This is Eq.~(\ref{eq:LEa}) in the main text.

The above equation can be generalized to finite temperatures, by distributing the operators at equal spacing around the thermal circle. Namely, let us define
\begin{equation}
    y^4=\frac{1}{Z}e^{-\beta H},
\end{equation}
and evaluate the OTOC as 
\begin{equation} \label{eq:regularization}
    F(t)=\Tr[y A^\dagger(t) y B^\dagger y A(t) y B].
\end{equation}
By absorbing the inverse temperature into the time evolution operator, the OTOC can be evaluated in the same manner as in the infinite temperature case, but in the complex time domain, $t-i\beta/4$. The averaged OTOC is translated to an infinite temperature average of LE in complex time, which is further reduced to a finite temperature average of the regular LE in real time. More precisely (using Eq.~(\ref{eq:LEb})),

\begin{equation}
\begin{aligned}
    \overline{F_{\beta}(t)} \approx & \left|\langle e^{i(H_B+V_1) (t+i\beta/4)}e^{-i(H_B+V_2) (t-i\beta/4)}\rangle\right|^2\\
    = & \frac{1}{Z_\beta} \frac{1}{d^2_B}\left|\Tr e^{i(H_B+V_1) t}e^{-\frac{\beta}{2} (H_B+O(V_1,V_2))}  e^{-i(H_B+V_2)t}\right|^2\\
    \approx & \frac{Z^2_{\beta/2}}{Z_\beta} \left|\langle e^{i(H_B+V_1) t}e^{-i(H_B+V_2) t}\rangle_{\beta/2}\right|^2 .
\end{aligned}
\end{equation}
Here $O(V_1,V_2)$ is an operator resulting from the Baker-Campbell-Hausdorff expansion that contains terms involving the perturbation operators $V_1$ and $V_2$. This term is time-independent and only contributes to a vanishing correction of the thermal state. Note that the effective inverse temperature for the LE is $\beta/2$. The factor $Z^2_{\beta/2}/Z_\beta$ appears because of the particular regularization (\ref{eq:regularization}). As a consequence, the OTOC at $t=0$ has a non-unit value, even for unitary operators $A$ and $B$, in contrast to the regular thermal average scheme, where the OTOC at $t=0$ is unit for unitary operators. In our numerical simulations, as a convention, OTOCs are always normalized by removing this prefactor. 

It is worth emphasizing that the Haar averaged OTOC is upper-bounded in the half strip $0<t$ and $-\frac{\beta}{4}\le \tau \le\frac{\beta}{4}$ for the complex time $t+i\tau$. The fact that the OTOC in this domain is bounded is crucial for the upper bound of the OTOC growth rate \cite{Maldacena2016-mb,Tsuji2018-td}. To see this fact, consider the averaged OTOC in complex time,
\begin{equation}
\begin{aligned}
    \overline{F_{\beta}(t)} \approx & \left|\langle e^{i(H_B+V_1) (t+i\tau+i\beta/4)}e^{-i(H_B+V_2) (t+i\tau-i\beta/4)}\rangle\right|^2\\
    = & \frac{1}{Z_\beta} \frac{1}{d^2_B}\left|\Tr e^{i(H_B+V_1) t}e^{-(\beta/4+\tau) (H_B+V_1)} e^{-(\beta/4-\tau) (H_B+V_2)} e^{-i(H_B+V_2)t}\right|^2,
\end{aligned}
\end{equation}
which is a bounded quantity for the effective positive temperature $\beta/4\pm \tau$.

\section{OTOC and LE for the inverted harmonic oscillator}\label{app:IHO}

In this section, we give the details for the derivation of the OTOC and LE for the inverted harmonic oscillator. 

\subsection{Wavefunction of the IHO}
The Hamiltonian of two coupled inverted harmonic oscillators (IHO) reads
\begin{equation}
    \sum_{i=1,2}\left(\frac{1}{2m_i}\hat{p}^2_i+\frac{m_i\omega_i^2}{2}x^2_i\right)+\delta \hat{x}_1\hat{x}_2.
\end{equation} 
Here, for convenience, the frequencies are written as imaginary numbers.
This Hamiltonian is reduced to the one of two uncoupled harmonic oscillators by the following transformation,
\begin{equation}
\begin{aligned}
    &x_1=\frac{1}{\sqrt{2}}\left(y_1+y_2\right),\\
    &x_2=\frac{1}{\sqrt{2}}\left(\eta y_1- \xi y_2\right),
\end{aligned}    
\end{equation}
where $\eta$ are $\xi$ are given by
\begin{equation}
\begin{aligned}
    &\eta=\left(\sqrt{D^2+4m_1/m_2}+D\right)/2,\\
    &\xi=\left(\sqrt{D^2+4m_1/m_2}-D\right)/2,\\
    &D=\left(\omega_1^2-\omega_2^2\right)m_1/\delta.
\end{aligned}    
\end{equation}
In the new coordinates, the effective masses and frequencies are given by
\begin{equation}
\begin{aligned}
    &\widetilde{m}_{1}=\frac{\left(\eta+\xi\right)^2}{2}\frac{m_1m_2}{m_1+\xi^2m_2}, \\
    &\widetilde{m}_{2}=\frac{\left(\eta+\xi\right)^2}{2}\frac{m_1m_2}{m_1+\eta^2m_2}, \\
    &\widetilde{m}_1\widetilde{\omega}^2_{1}=\left(m_1\omega^2_1+\eta^2m_2\omega^2_2+\delta\eta\right),\\
    &\widetilde{m}_2\widetilde{\omega}^2_{2}=\left(m_1\omega^2_1+\xi^2m_2\omega^2_2+\delta\xi\right).\\
\end{aligned}    
\end{equation}

The regularized thermal state evolved in the OTOC and LE is replaced with a pure product state (up to a normalization factor) $|\psi(x_1)\rangle|\psi(x_2)\rangle=e^{-x_1/m_2}e^{-x_2/m_1}$. The choice of this wavefunction is not physically essential for the OTOC and LE. It is however important from a practical point of view. This particular form of wavefunction is separable in the new coordinates $y_1$ and $y_2$, which makes solving the time-dependent wavefunction much easier.

For a given initial state $\psi(x)$ of a single IHO with mass $m$ and frequency $\omega$, the time-dependent wavefunction can be expressed by a path integral
\begin{equation}
    \psi(x,t)=\int dx'\ K(x,x',t,0)\psi(x',0), 
\end{equation}
where the propagator is
\begin{equation}
\begin{aligned}
        K(x&,x',t,0)=\sqrt{\frac{m\omega}{2\pi i \sin(\omega t)}}\\
        &\times \exp{[\frac{im\omega}{2\sin{\omega t}}\left((x^2+x'^2)\cos{\omega t}-2xx'\right)]}.
\end{aligned}
\end{equation}
With these toolkits, the wavefunction $\psi(x_1,x_2,t)$ (which is involved in the OTOC evaluation; see the following section) can be expressed in the new coordinates using the single IHO propagator and then transfer back to the original coordinates.

\subsection{Haar averaged OTOC for the IHO}
In this section, we compute the Haar averaged OTOC for the IHO:
\begin{equation}\label{eq:app_ootciho}
    \overline{F(t)}=\int_{Haar} dAdB \Tr[ A^\dagger(t)\rho^{\frac{1}{4}} B^\dagger\rho^{\frac{1}{4}} A(t) \rho^{\frac{1}{4}}B \rho^{\frac{1}{4}}].
\end{equation}
The operators $A$ and $B$ to be averaged are unitary operators on the first and second IHO, respectively. Since a termal state of the IHO is not well defined, we replace the regularized thermal state $\rho_\beta$ with a pure state of the coupled IHO, i.e., $\rho(x_1,x_2;x_1',x_2')=\psi^*(x_1,x_2)\psi(x_1',x_2')=\rho^{\frac{1}{4}}(x_1,x_2;x_1',x_2')$.
Using the property of the Haar measure described in Appendix \ref{app:Haar}, the integral can be computed directly:
\begin{equation}
\begin{aligned}
        &\int_{Haar} dAdB\  A^\dagger(t)\rho B^\dagger\rho A(t) \rho B \rho\\
        =&\int dAdB\  A^\dagger e^{-iHt}\rho B^\dagger\rho e^{iHt} A e^{-iHt} \rho B \rho e^{iHt}\\
        =&\int dB\ \Tr_A[e^{-iHt}\rho B^\dagger\rho e^{iHt}]\times e^{-iHt} \rho B \rho e^{iHt}\\
        =&\int dB\ \int dx_1\langle x_1|e^{-iHt}\rho B^\dagger\rho e^{iHt}|x_1\rangle e^{-iHt} \rho B \rho e^{iHt}\\
        =&\int dx_1 \langle x_1|e^{-iHt}\rho \times \Tr_B \left(\rho e^{iHt}|x_1\rangle e^{-iHt} \rho\right) \times \rho e^{iHt}\\
        =&\int dx_1 \langle x_1|e^{-iHt}\rho \langle x_2|\rho e^{iHt}|x_1\rangle e^{-iHt} \rho |x_2\rangle \times \rho e^{iHt}\\
        =&\int dx_1dx_2 \underbrace{\langle x_1|e^{-iHt}\rho}_{a} \underbrace{\langle x_2|\rho e^{iHt}|x_1\rangle}_{b} \underbrace{e^{-iHt} \rho |x_2\rangle}_{c} \underbrace{\rho e^{iHt}}_{d}\\
\end{aligned}
\end{equation}
Using the expression for the density operator, i.e.,
\begin{equation}
\begin{aligned}
     e^{-iHt}\rho=\int dx'_1dx_1''dx'_2dx_2''\ \ \psi^*(x_1',x_2',t)\psi(x_1'',x_2'')|x_1',x_2'\rangle\langle x_1'',x_2''|, 
\end{aligned}
\end{equation}
the four terms in the above Haar integral can be evaluated separately as
\begin{equation}
\begin{aligned}
        a=&\int dx_1''dx'_2dx_2''\psi^*(x_1,x_2',t)\psi(x_1'',x_2'')|x_2'\rangle\langle x_1'',x_2''|,\\
        b=&\int dx_1''dx'_2\psi(x_1,x_2',t)\psi^*(x_1'',x_2)|x_1''\rangle\langle x_2'|,\\
        c=&\int dx'_1dx_1''dx'_2\psi^*(x_1',x_2',t)\psi(x_1'',x_2)|x_1',x_2'\rangle\langle x_1''|,\\
       d= & \int dx'_1dx_1''dx'_2dx_2''\ \ 
        \psi(x_1',x_2',t)\psi^*(x_1'',x_2'')|x_1'',x_2''\rangle\langle x_1',x_2'|.
\end{aligned}
\end{equation}
Besides the above Haar integral, the OTOC Eq.~(\ref{eq:app_ootciho}) involves a total trace, i.e.,
\begin{equation}
    \Tr\bullet = \int dx_1dx_2 \langle x_1|\langle x_2| \bullet |x_1\rangle|x_2\rangle.
\end{equation}
After evaluating the trace and Haar integral, the OTOC finally reads
\begin{equation}
\begin{aligned}
        \overline{F(t)}&=\int dx_1dx_1'dx_2dx_2' \ \
        \psi(x_1,x_2,t)\psi^*(x_1',x_2,t)\psi(x_1',x_2',t)\psi^*(x_1,x_2',t)\\
        &=\int dx_2dx_2'\  \rho_2(x_2',x_2)\rho_2(x_2',x_2)\\
        &=\int dx_2dx_2'\  \rho^*_2(x_2,x_2')\rho_2(x_2',x_2)\\
        &=\int dx_2 \rho_2^2(x_2)=\Tr(\rho_2^2).
\end{aligned}
\end{equation}
Here, $\rho_2$ is the reduce density matrix of the second oscillator. The above equation is an exact result. In this particular setting, where the regularized thermal state is replaced with a pure state, the OTOC is precisely equal to the purity of one subsystem. The purity can be computed exactly as well, using the procedure described in the previous section.

\subsection{\emph{Baker-Campbell-Hausdorff} expansion of the LE}
In this section we derive the second order expansion with respect to the perturbation strength $\delta$ for the Loschmidt echo of the type
\begin{equation}
    M(t)=|\langle\psi|e^{i(H-\delta\hat{x})t}e^{-i(H+\delta\hat{x})t}|\psi\rangle|^2,
\end{equation}
where $H$ is the Hamiltonian of a single IHO
\begin{equation}
    H=\frac{1}{2m}\hat{p}^2+\frac{m\omega^2}{2}\hat{x}^2.
\end{equation}
Here $\omega$ is a pure imaginary number.
Treating $H_0\equiv H-\delta \hat{x}$ as the ``unperturbed'' Hamiltonian, the echo operator reads
\begin{equation}\label{eq:echou}
    \hat{U}_t\equiv e^{i(H-\delta\hat{x})t}e^{-i(H+\delta\hat{x})t}=e^{iH_0t}e^{-i(H_0+\delta2\hat{x})t}.
\end{equation}
It can be expanded to the second order in $\delta$ using the \emph{Baker-Campbell-Hausdorff} (BCH) formula \cite{Gorin2006-ro}:
\begin{equation}
    \hat{U}_t=\exp{\left[-i\left(\Sigma(t)\delta+\frac{1}{2}\Gamma(t)\delta^2+\dots\right)\right]},
\end{equation}
where $\Sigma(t)$ and $\Gamma(t)$ are determined by the Heisenberg evolution of the perturbation operator, i.e., $2\hat{x}(t)\equiv e^{iH_0t}2\hat{x}e^{-iH_0t}$,
\begin{equation}
\begin{aligned}
    \Sigma(t)&=\int_0^tdt'\ 2\hat{x}(t')\\
    \Gamma(t)&=i\int_0^t dt'\int_{t'}^tdt''\ [2\hat{x}(t'),2\hat{x}(t'')].
\end{aligned}
\end{equation}

To solve the Heisenberg evolution of $2\hat{x}$ under Hamiltonian $H_0$, we perform a coordinate transformation $y=x+2\delta/(m\omega^2)$. In the new coordinate the Hamiltonian reads
\begin{equation}
    H_0=\frac{1}{2m}\hat{p}^2+\frac{m\omega^2}{2}\hat{y}^2-\frac{2\delta^2}{m\omega^2}.
\end{equation}
This is a shifted harmonic oscillator Hamiltonian, for which the Heisenberg evolution of the position operator has been computed \cite{Hashimoto2017-ug}:
\begin{equation}
    \hat{y}(t)=\hat{y}\cos{\omega t}+\frac{1}{m\omega}\hat{p}\sin{\omega t},
\end{equation}
which gives
\begin{equation}
\begin{aligned}
    \hat{x}(t)=\hat{x}\cos{\omega t}+\frac{1}{m\omega}\hat{p}\sin{\omega t}+\frac{2\delta}{m\omega^2}\left(\cos{\omega t}-1\right).
\end{aligned}
\end{equation}
The last term in the above equation will be dropped since it contributes to higher than second-order in $\delta$ for the LE. $\Sigma$ and $\Gamma$ are thus given by
\begin{equation}
\begin{aligned}
        \Sigma(t)=&\frac{2}{\omega}\hat{x}\sin{\omega t}-\frac{2}{m\omega^2}\hat{p}\cos{\omega t}+\frac{2}{m\omega^2}\hat{p},\\
        \Gamma(t)=&\frac{4}{m\omega^3}\sin{\omega t}-\frac{4}{m\omega^2}t.
\end{aligned}
\end{equation}

In the following, we perform a Taylor expansion of the LE to second order in $\delta$. The second order expansion for the average of the echo operator (\ref{eq:echou}) is given by
\begin{equation}
\begin{aligned}
    \langle\psi|\hat{U}_t|\psi\rangle=1-i\left(\overline{\Sigma(t)}\delta+\frac{1}{2}\overline{\Gamma(t)}\delta^2\right)-\frac{1}{2}\overline{\Sigma^2(t)}\delta^2.
\end{aligned}
\end{equation}
Note that both $\Sigma(t)$ and $\Gamma(t)$ are real, there are only non-vanishing two second-order terms kept in the magnitude-square of the above quantity, which gives the final expression of the LE:
\begin{equation}
\begin{aligned}
        M(t)&=|\langle\psi|\hat{U}_t|\psi\rangle|^2=1-\left(\overline{\Sigma^2(t)}+\overline{\Sigma(t)}^2\right)\delta^2.
\end{aligned}
\end{equation}

As defined in the previous sections, the pure state for the OTOC average is chosen as a Gaussian form with zero mean. In this case only the first term in $\Sigma(t)$ has a non-zero contribution to the LE $M(t)$, i.e.,
\begin{equation}
\begin{aligned}
      M(t)=&1-\frac{\delta^2}{\omega^2}\sin^2{\omega t}\langle\psi|(2\hat{x})^2|\psi\rangle\\
      =&1-\delta^2\frac{4\overline{\hat{x}^2}}{\omega^2}\sin^2{\omega t}.
\end{aligned}
\end{equation}

\section{Finite size study of the SYK model.}\label{app:SYK}

\begin{figure}[H]
\centering
\includegraphics[width=0.8\columnwidth]{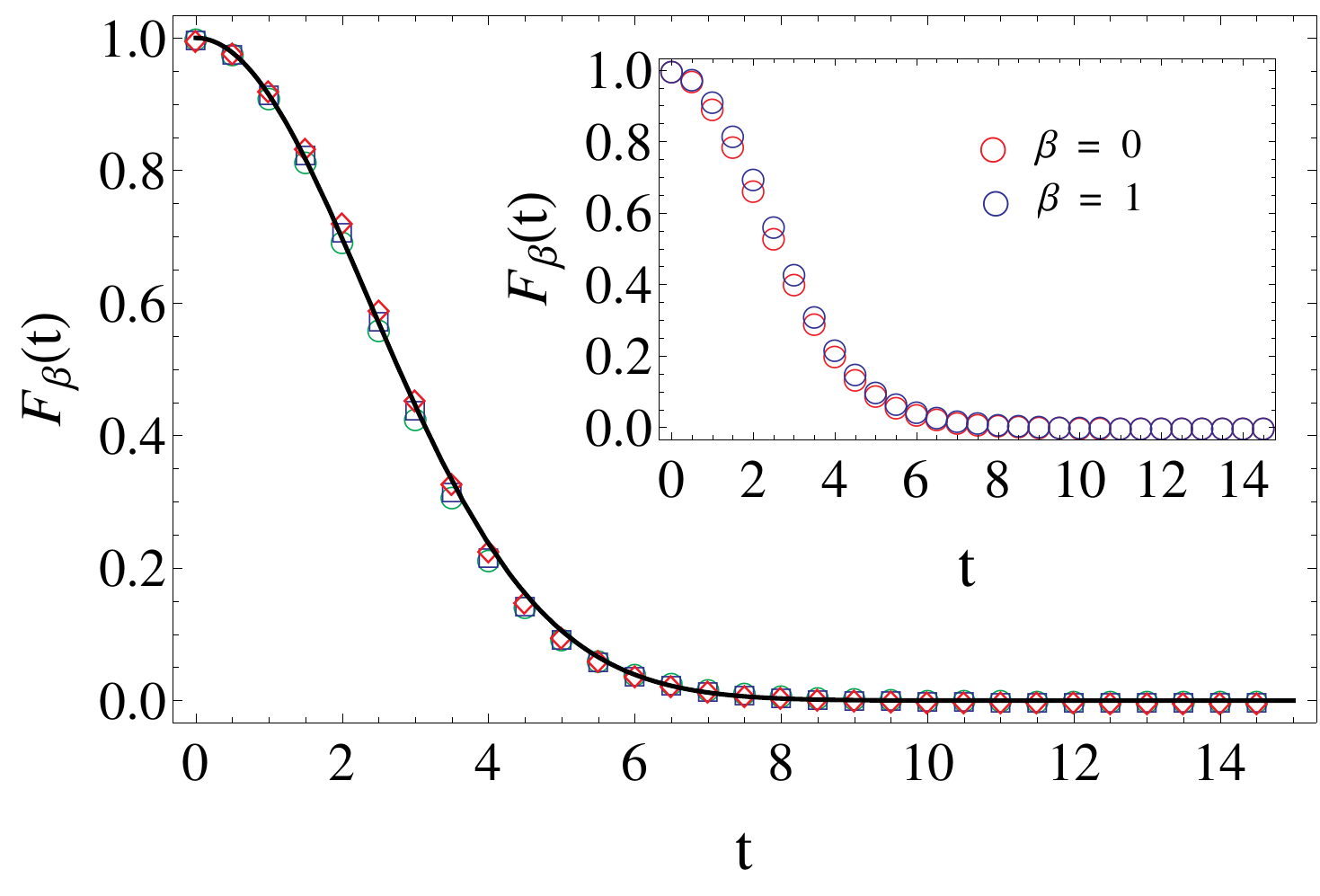}
\caption{\label{fig:syk}
OTOC for the fermionic SYK model at infinite temperature. The coupling strength $J_{i,j;k,l}$ has a unit variance. Green circles, blue squares and red diamonds correspond to $N=15, 16, 17$, respectively. The black solid curve is the best fit to a Gaussian decay. Inset: OTOC for 16 fermions at the same coupling strength at different temperatures.}
\end{figure}

To further test our theory, we present numerical study of the fermionic version of the SYK model proposed in \cite{Fu2016-dg}. The Hamiltonian of the SYK model reads
\begin{equation}
    H=\frac{1}{(2N)^{3/2}}\sum_{i,j,k,l=1}^N J_{i,j;k,l}c^\dagger_ic^\dagger_jc_kc_l,
\end{equation}
where $J_{i,j;k,l}$ are complex Gaussian random couplings with zero mean obeying certain symmetries. $c_i$ and $c_i^\dagger$ are fermionic annihilation and creation operators at site~$i$. We compute the OTOC for operators $c_i^\dagger+c_i$ on two distinct sites. Clear Gaussian decay with very weak temperature dependence has been observed (Fig.~\ref{fig:syk}), which agrees with previous numerical studies \cite{Fu2016-dg,Hosur2016-gb} while contradicting the expected exponential decay with upper bound $2\pi/\beta$ \cite{Maldacena2016-mb,Tsuji2018-td}. 

On the other hand, this observation fits into the theory of the present work: The OTOC decay rate is governed by the coupling strength between the target subsystem $S_A$, $S_B$ and the rest of the system. In a finite system, the coupling might be too large for the decay to be in the exponential regime. The decay rate extracted from our numerical simulation is close to the bandwidth (the width of the spectral density) of the SYK Hamiltonian, which indicates that the decay is in the Gaussian regime. 

\begin{figure}[H]
\centering
\includegraphics[width=0.8\columnwidth]{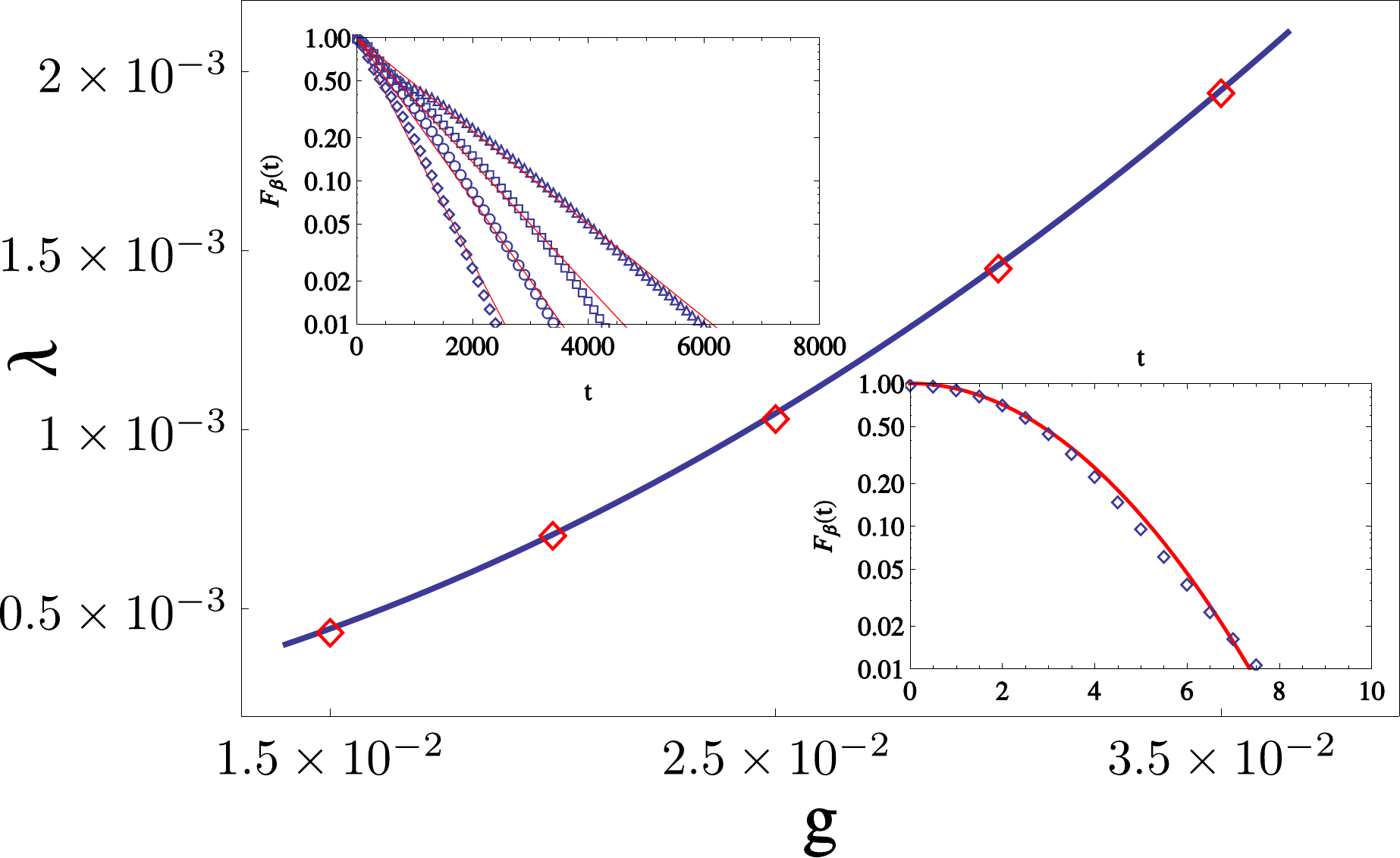}
\caption{
Exponential decay rate $\lambda$ v.s. $g$ factor. The blue solid line is the best fit to the quadratic form. The decay rates are extracted from the numerical simulation of the OTOC evolution for the fermionic SYK model of 16 fermions with manually decreased couplings (by a factor $0<g<1$) between the target subsystem and the rest of the whole system, see text for details. Inverse temperature is fixed at zero. The data are obtained by averaging over 50 realizations. The top inset shows typical decay curves at decreased coupling strength. The triangles, squares, circles and diamonds correspond to $g=0.02, 0.025, 0.03$ and $0.035$, respectively. The bottom inset shows the Gaussian decay without decreasing the coupling strength ($g=1$).}
\label{fig:deform}
\end{figure}

Since the relative strength of the coupling between the two subsystems $S_A$ and $S_B$ compared with the total Hamiltonian decreases with the system size, it is then expected that the OTOC decay would drop into the exponential regime in the large-$N$ limit. Due to the limited numerical capacity, we did not observe exponential decay up to 17 fermions, which is the largest system we are able to simulate numerically. However, we are able to observe exponential decay by manually adjusting the coupling strength. It is done by decreasing the couplings $J_{i,j;k,l}$ that involve the subsystem $S_A$ and $S_B$ by a factor $0<g<1$, while keeping the coupling in the rest of the system unchanged.

Figure~\ref{fig:deform} shows the exponential OTOC decay for the deformed coupling strength. The decay rate also admits quadratic dependence on the coupling factor $g$, which satisfies the Fermi's golden rule prediction for the exponential decay of the LE. 

\clearpage

\end{document}